\documentclass[aps,reprint,amsmath,twocolumn,amssymb,floatfix]{revtex4-1}
\usepackage{natbib}
\usepackage{graphicx}
\usepackage{bm}
\usepackage{color,soul}
\usepackage{hyperref}
\usepackage{float}
\usepackage{csquotes}
\usepackage{color,soul}
\hypersetup{colorlinks=true, urlcolor=blue, citecolor=blue}
\usepackage{braket}
\usepackage{amsmath}
\usepackage{rotating}
\DeclareUnicodeCharacter{2005}{\hspace{0.25em}}

\begin{document}
\preprint{APS}
\title{ Electronic structure and optoelectronic properties of halide double perovskites:  Fundamental insights and design of a theoretical workflow}
\author{Mayank Gupta}
\affiliation{Condensed Matter Theory and Computational Lab, Department of Physics, Indian Institute of Technology Madras, Chennai  600036, India}
\affiliation{Center for Atomistic Modelling and Materials Design, Indian Institute of Technology Madras, Chennai  600036, India}
\author{Susmita Jana}
\affiliation{Condensed Matter Theory and Computational Lab, Department of Physics, Indian Institute of Technology Madras, Chennai  600036, India}
\affiliation{Center for Atomistic Modelling and Materials Design, Indian Institute of Technology Madras, Chennai  600036, India}
\author{B. R. K. Nanda}
\email{nandab@iitm.ac.in}
\affiliation{Condensed Matter Theory and Computational Lab, Department of Physics, Indian Institute of Technology Madras, Chennai  600036, India}
\affiliation{Center for Atomistic Modelling and Materials Design, Indian Institute of Technology Madras, Chennai  600036, India}
\date{\today}
\begin{abstract}
Like single perovskites, halide double perovskites (HDP) have truly emerged as efficient optoelectronic materials since they display superior stability and are free of toxicity. However, challenges still exist due to either wide and indirect bandgaps or parity-forbidden transitions in many of them. The lack of understanding in chemical bonding and the formation of parity-driven valence and conduction band edge states have hindered the design of optoelectronically efficient HDPs. In this study, we have developed a theoretical workflow using  a multi-integrated approach involving ab-initio density functional theory (DFT) calculations, model Hamiltonian studies, and molecular orbital picture leading to momentum matrix element (MME) estimation. This workflow gives us detailed insight into chemical bonding and parity-driven optical transition between edge states. In the process, we have developed a band-projected molecular orbital picture (B-MOP) connecting free atomic orbital states obtained at the Hartree-Fock level and orbital-resolved DFT bands. From the B-MOP, we show that the nearest neighbor cation-anion interaction determines the position of atom-resolved band states, while the second neighbor cation-cation interactions determine the shape and width of band dispersion and, thereby, MME. The latter is critical to quantify the optical absorption coefficient. Considering both B-MOP and MME, we demonstrate a mechanism of tailoring bandgap and optical absorptions through chemical doping at the cation sites. Furthermore, the cause of bandgap bowing, a common occurrence in doped HDPs, is explained by ascribing it to chemical effect and structural distortion.
\end{abstract}
\maketitle
\section{Introduction:}
In the last couple of decades, organic and inorganic halide single perovskites (HSPs) of the formula ABX$_3$ (e.g. CsPbI$_3$) have gained enormous research attention as they demonstrate promising optoelectronic  properties \cite{ChemRev, NanoLett_aron}, solar cell applications \cite{Tsutomu-JACS}, non-trivial topological quantum phases which bring the dimension of orbitronics \cite{mayank-2022, Kepenekian-2015} and topotronics \cite{jin-PRB,ravi-2020,kore-2020}.  At the same time, there appears to be a large number of disadvantages associated with this class of compounds. The most significant one is the lack of stability on prolonged exposure to light and heat. As most of the promising HSPs are lead (Pb) based, toxicity remains another concern. The halide double perovskites (HDPs) are emerging as an alternate class of compounds which to some extent, overcome the aforementioned disadvantages. HDP has a general formula of A$_2$BB$^\prime$X$_6$ where A is a monovalent cation of Group-I, B and B$^\prime$ are metals with +1 (K, Na, Ag, Au, Cu, In, Tl) and +3 (Bi, Sb, In, Tl) oxidation states, and X is a halide. Most commonly, in HDPs, A-site is Cs, and Cl and Br are considered halogen sites. \par
Compared to the HSPs, HDPs are in general more stable \cite{BIBI2021123} and environmental friendly. They create a large chemical configurational space, and therefore this family is capable of exhibiting diverse electronic structures and, in turn, are suitable for a wide range of applications. These include photovoltaic solar cells \cite{Zhang2022}, photodetectors \cite{JMCC_Tong}, photocatalysis \cite{Muscarella-2022, Li-2022}, CO$_2$ reduction \cite{co2}, spintronics \cite{ning-2020},  X-ray detectors \cite{Zhu2020,ZHANG2020}, water splitting \cite{poonam}, etc. The HDPs are also being actively examined as solar cell absorbers. However, the issue of indirect bandgap in some of them and parity forbidden transition in others \cite{DP1,DP2,DP3,DP4,DP5} are the bottleneck which needs to be addressed. For example, Cs$_2$AgBiBr$_6$ and Cs$_2$AgSbCl$_6$ possess indirect bandgap. On the other hand, though Cs$_2$AgInCl$_6$ exhibits a direct band gap of 3.23 eV, parity forbidden transition at $\Gamma$, leads to very weak optical absorption near the band gap. The next optical transition in this system happens at $\sim$ 4.46 eV, which is much higher for an ideal solar cell material \cite{Zhou2017}. In a recent study, it has been revealed that, for B = In$^+$, TI$^+$ and B$^\prime$ = Sb$^{3+}$, Bi$^{3+}$, HDPs show favorable optical absorption suitable for thin-film solar cell applications\cite{AgTlCl}. Unfortunately, Tl$^+$ is toxic, and In$^+$ tends to be unstable against oxidation and form mixed-valence compounds with distorted and complex crystal structures \cite{Xiao2017}. \par
Despite a few disadvantages, HDPs have attracted considerable attention due to their simple, robust, and easy synthesis process. Since the valence band and conduction band edges are formed out of the covalent hybridization among the orbitals of metal cations (B, B$^\prime$) and halide anions (X), cationic and anionic mixing naturally becomes an effective strategy to manipulate the electronic properties and optical behavior. Taking into account these advantages, many design principles have been proposed experimentally \cite{Prasanna-2017} and theoretically \cite{Zhang-2017, mark-2019, Joule-2018} to modify the electronic structure so as to achieve better optoelectronic performances.\par
Recent studies \cite{Appadurai2021, aswani-2019}  have shown that, Cs$_2$AgIn$_x$Bi$_{1-x}$Cl$_6$ and Cs$_2$AgBi$_x$Sb$_{1-x}$Cl$_6$ produce high photoluminescence for the range 0.8 $\leq$ x $\leq$ 0.9. The reasons are attributed to the manipulation of bandgap and parity. In another study, Athrey and Sudakar \cite{athrey1,athrey2} have experimentally demonstrated distortion drive nonlinear bandgap variation and self-trapped exciton (STE) emission on the cationic intermix systems Cs$_2$(Ag, Na)BiCl$_6$. Interestingly,  the anionic intermixing (Cs$_2$AgBiBr$_{6-x}$Cl$_x$) results in linear bandgap variation \cite{Athrey4}. In a combined theoretical and experimental study, Slavney et al. \cite{Slavney-2017} reported a change in the bandgap from 1.95 to 1.4 eV in MA$_2$AgBiBr$_6$ by Tl doping at Ag and Bi site, which is close to the ideal bandgap for photovoltaic applications. Many other studies are carried out to demonstrate the tuning of optical properties in HDPs by a similar cation intermixing approach \cite{Du-2017, karmakar-2018,trans-2017,Zhang2022}. However, these isolated investigations with limited scopes do not reveal the universal mechanism that alters the electronic structure at the vicinity of the Fermi level. Hence, there is a lack of guiding principles which can be utilized to design HDPs for electronic applications through controlled cationic and anionic intermixing. Developing materials design workflow has become necessary as HDPs are now being intensely investigated in search of stable and highly efficient solar cell materials.\par
\begin{figure}[ht]
\centering
\includegraphics[angle=-0.0,scale=0.22]{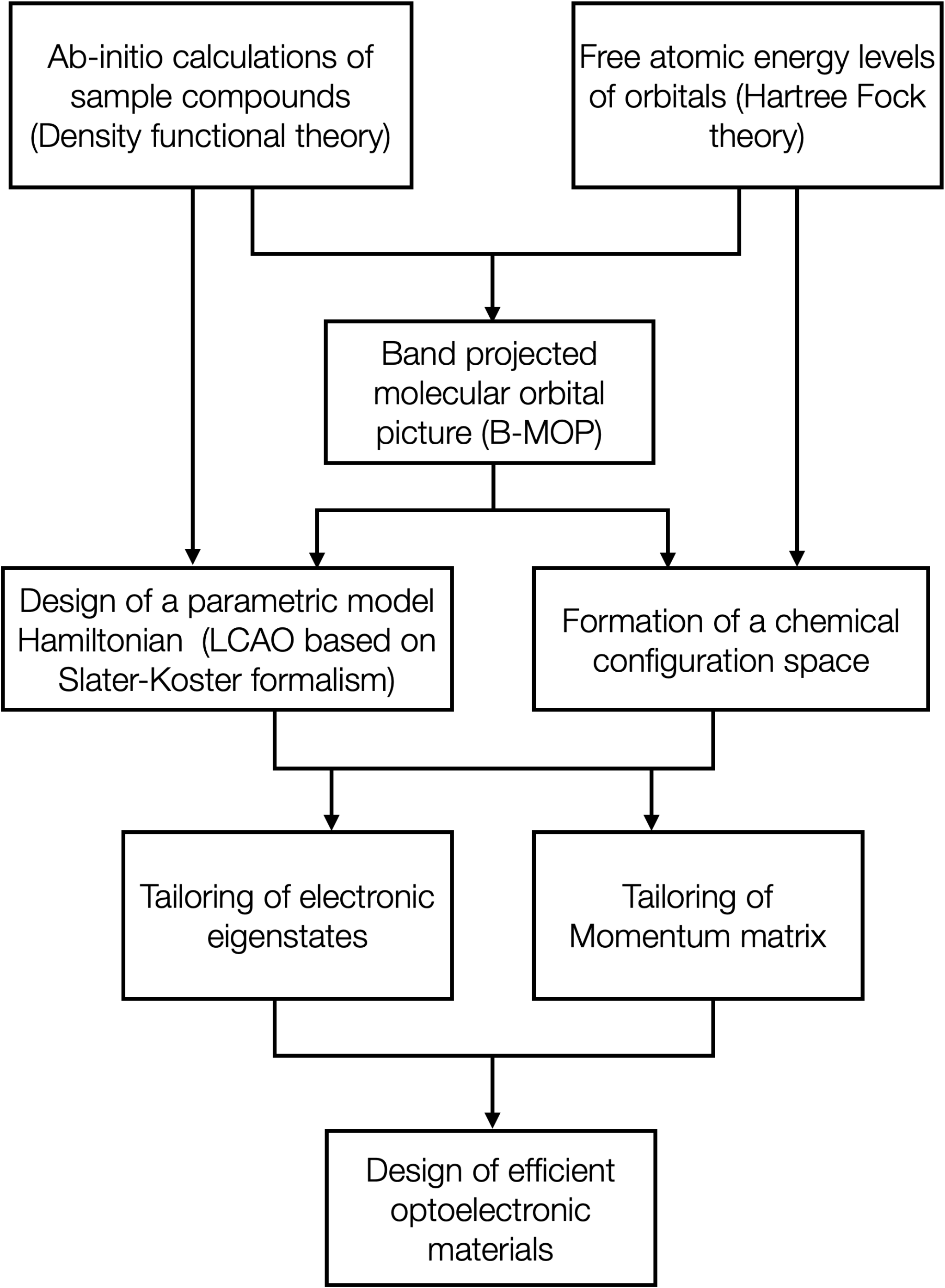}
\caption{A schematic summarizing the design principle to calculate and predict the efficient optoelectronic properties of HDPs.}
\label{Fig1}
\end{figure}

In this study, by considering a set of prototype compounds Cs$_2$BB$^\prime$Cl$_6$ (B = Ag, Na, In, Tl and B$^\prime$ = In, Tl, Bi, Sb), we develop a theoretical workflow to establish the relationship between cationic and anionic intermixing and the electronic structure, as well as, the optical absorption in the HDPs. The theoretical workflow, schematically illustrated in Fig. \ref{Fig1}, is based on density functional theory (DFT) calculations, an elegant Slater-Koster formalized tight-binding (SK-TB) model, and band projected molecular orbital picture (B-MOP). The optical absorption study is  carried out by calculating the momentum matrix elements (MME), which are the outcomes of the solution to model Hamiltonian. Through the workflow, we understand the chemical bonding and parity-driven optical transition between the edge states. With the aid of B-MOP, we infer that the nearest neighbor cation-anion interaction determines the position of atom-resolved band states. On the other hand, second neighbor cation-cation interactions determine shape and width of the band dispersion and, hence, the MMEs. The imaginary part of the dielectric constant and in turn the optical absorptions are calculated using the MMEs. With the aid of both B-MOPs and MMEs, we demonstrate how chemical doping at the cation site can tailor the bandgap and optical absorption. As a byproduct, we demonstrate how the chemical effect and structural distortion together cause bandgap bowing, a common occurrence in doped HDPs.
\section{Designing Approach and Computational Details}
\begin{figure}
\centering
\includegraphics[scale=0.38]{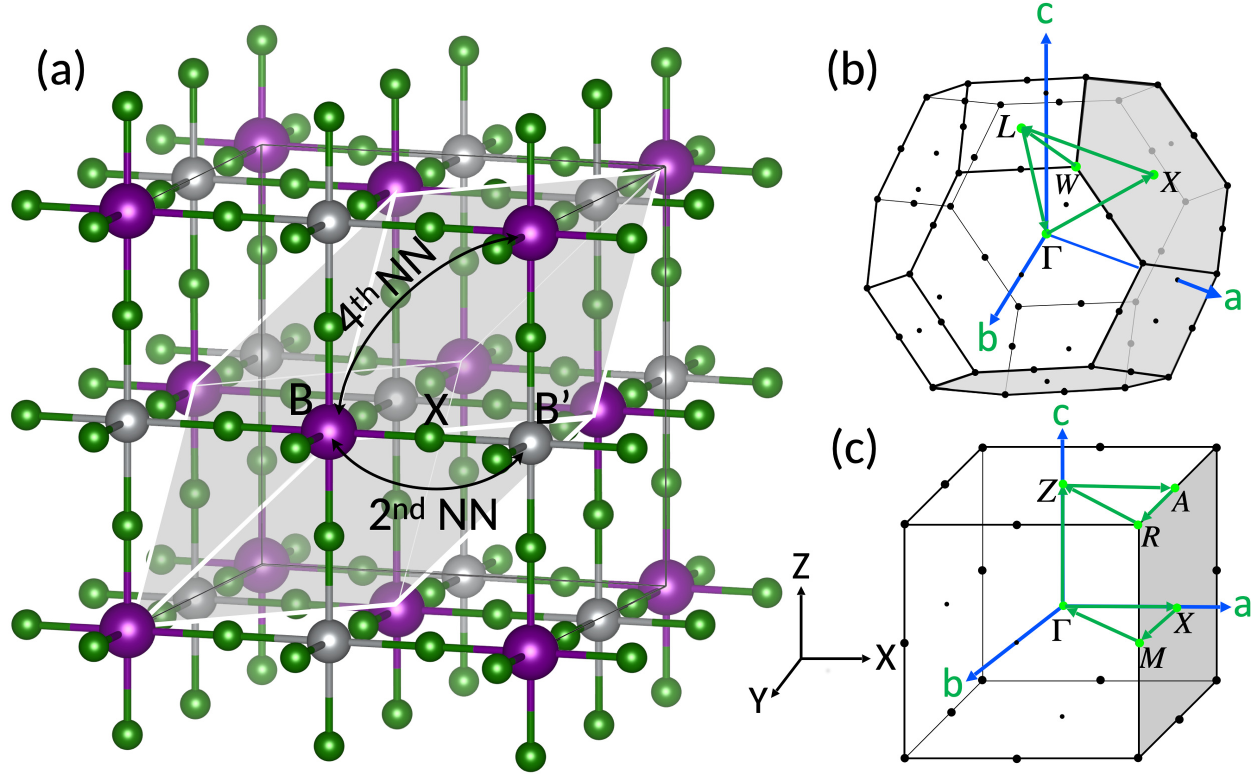}
\caption{(a) The rhombohedral (one formula unit) unit cell (shown in shaded gray color) inside a conventional (four formula unit) face-centered cubic cell  of the halide double perovskite A$_2$BB$^\prime$X$_6$. The 2$^{nd}$ nearest-neighbor (NN) electron-electron hopping interactions take place between B and B$^\prime$ cations and are mediated by halogen X anions. The 4$^{th}$-NN interactions occur between the same metal cations $i.e.$, B to B and B$^\prime$ to B$^\prime$. (b and c) The first Brillouin zone for the rhombohedral and conventional unit cells, respectively. The $k$-path used for band structure calculations is shown in green color.}
\label{Fig2}
\end{figure}

We will first briefly discuss the crystal structure of HDPs. As shown in Fig. \ref{Fig2} (a), it has a single unit rhombohedral primitive unit cell (Fm$\bar{3}$m) with two organic or inorganic monovalent A cations, one monovalent B, one trivalent B$^\prime$ cations, and six halogen anions. A conventional crystal structure of HDP is a cubic unit cell and  contains four formula units. The salient feature of the crystal structure is the  presence of BX$_6$ and B$^\prime$X$_6$ octahedra which are alternately arranged and connected with corner-sharing X-anions in all three directions. The A$^+$ cations occupy the cuboctahedral cavities positions. \par
The approach to design the theoretical workflow is summarized in the flowchart shown in Fig. \ref{Fig1}.  Hartree-Fock calculations on sample free atoms A, B, B$^\prime$, and X provide the free atomic orbital energy levels. This, in combination with the DFT calculated band structure, establishes B-MOP describing the possible chemical bondings of the prototype compounds. The B-MOPs enable us to design a parametric tight-binding model Hamiltonian and construction of chemical configuration space. The variation in the parameter and configuration can contribute towards searching for desired electronic structure and optical absorption deterministic momentum matrix so as to maximize optoelectronic efficiencies. Each component of the flowchart is further described in detail in the remaining part of the paper.\par
The DFT electronic structure calculations are performed on a set of HDPs (see Table \ref{T1}) using full-potential linearized augmented plane-wave (FP-LAPW) method as implemented in the WIEN2k simulation tool \cite{Wien2k}. For structural relaxation, we have used pseudopotential-based Vienna ab-initio Simulation Package (VASP)\cite{vasp} within the framework of the projector-augmented waves (PAW) method. Relaxations are performed via the conjugate gradient algorithm until the residual force in each atom is $<$ 0.01 eV/\AA. A $k$-mesh of size 6 $\times$ 6 $\times$ 6 is used for the Brillouin zone (BZ) sampling, PBE generalized gradient approximation (GGA) \cite{PBE1,PBE2} is employed for the exchange-correlation functional with the energy cutoff 400 eV, and the convergence criterion for total energy is set to 10$^{-6}$ eV. The lattice constants of HDPs after relaxation are provided in Table \ref{T1}. \par
The GGA-PBE functional underestimates the bandgap as compared to the experimental bandgap, and hence the GGA-PBE along with the modified Becke–Johnson (mBJ) \cite{mBJ} potential is used to calculate the electronic band structure of HDPs. The results are in good agreement with the experimental bandgap (see Table \ref{T1}). The number of plane waves is determined by setting $R_{MT}K_{MAX}$ to 7.0 for all the compounds. The BZ integration is carried out with a Monkhorst–Pack grid with the k-mesh size of 8 $\times$ 8 $\times$ 8 (yielding 35 irreducible points). The calculations include spin-orbit coupling (SOC) effect. For the  model Hamiltonian studies, we have developed a few codes using MATLAB \cite{MATLAB}, and the package is made available online \cite{MayankGithub}.\par
\begin{table*}
\centering
\caption {DFT obtained GGA+mBJ bandgaps for various halide double perovskites along with the nature of bandgap, orbital compositions of valence and conduction bands, and valence electron counts (VEC).}
\begin{tabular}{|c|c|c|c|c|c|c|c|c|} 
\hline
Crystal formula &	Lattice Constant(\AA) &	Lattice Constant(\AA) & Bandgap & Bandgap &	Nature of &	VBM  &	CBM  &	VEC \\
& (GGA relaxed) & (Experimental) &	GGA-mBJ(eV) & exp (eV) &bandgap&&& \\
\hline
Cs$_2$AgBiBr$_6$ &	11.465	&11.271 \cite{Zhang2022,athrey3}&	1.90  &2.19 \cite{Zhang2022}&	Indirect &	Ag-$e_g$+Bi-$s$  &	Bi-$p$ &	48\\
Cs$_2$AgBiCl$_6$ &	10.936	&10.760 \cite{athrey1}&	2.98  & 2.65 \cite{athrey1}&	Indirect &	Ag-$e_g$+Bi-$s$ &	Bi-$p$ &	48\\
Cs$_2$AgSbCl$_6$ &	10.809	&10.699 \cite{karmakar-2018} &	2.41 & 2.57 \cite{karmakar-2018}&	Indirect &	Ag-$e_g$+Sb-$s$  &	Sb-$p$ &	48\\
Cs$_2$AgInCl$_6$ &	10.560	&10.480 \cite{Zhang2022}&	3.28 & 3.57 \cite{AgInCl_gap}&	Direct &	Ag-$e_g$ &	In-$s$ &	46\\
Cs$_2$AgTlCl$_6$ &	10.784	&10.559 \cite{AgTlCl}& 	0.64 &1.96 \cite{AgTlCl}&	Direct &	Ag-$e_g$ &	Tl-$s$ &	46\\
Cs$_2$AgAsCl$_6$ &	9.14	& -  & 2.30 & -&	Indirect &	Ag-$e_g$+As-$s$  &	As-$p$ &	48\\
Cs$_2$AgGaCl$_6$ &	9.00	& -  & 2.89 & -&	Direct &	Ag-$e_g$ &	Ga-$s$ &	46\\
Cs$_2$CuBiCl$_6$ &	10.60	& -   &0.82 & 1.57 \cite{NEELU2023114250}&	Indirect &	Cu-$e_g$+Bi-$s$ &	Bi-$p$ &	48\\
Cs$_2$CuInCl$_6$ &	10.34	& -   &0.25 & -&	Direct &	Cu-$e_g$ &	In-$s$ &	46\\

Cs$_2$InBiCl$_6$ &	11.345	&-&	0.53 &-&	Direct &	In-$s$ & 	Bi-$p$ &	40\\
Cs$_2$InSbCl$_6$ &	11.212	&-&	0.64 &-&	Direct &	In-$s$ &	Sb-$p$ &	40\\
Cs$_2$TlBiCl$_6$ &	11.547	&-&	2.09 &-&	Direct &	Tl-$s$ &	Bi-$p$ &	40\\
Cs$_2$TlSbCl$_6$ &	11.420	&-&	2.01 &-&	Direct &	Tl-$s$ &	Sb-$p$ &	40\\
Cs$_2$TlBiBr$_6$ &	12.069	&-&	1.34 &-&	Direct &	Tl-$s$ &	Bi-$p$ &	40\\
Cs$_2$TlSbBr$_6$ &	11.944	&-&	1.27 &-&	Direct &	Tl-$s$ &	Sb-$p$ &	40\\
Cs$_2$NaBiCl$_6$ &	11.026	&10.833 \cite{athrey2}&	4.15 & 3.0 \cite{athrey1}&	Indirect &	Bi-$s$ &	Bi-$p$ &	38\\
Cs$_2$NaSbCl$_6$ &	10.930	&-&	3.99 &-&	Indirect &	Sb-$s$ &	Sb-$p$ &	38\\
Cs$_2$NaInCl$_6$ &	10.730	&10.533 \cite{NaInCl}&	5.37 & 4.15 \cite{NaInCl-gap}&	Direct &	Cl-$p$ &	In-$s$ &	36\\
Cs$_2$KInCl$_6$  &  11.156  &10.770 \cite{CsKInCl}&   5.89 &-& Direct &  Cl-$p$  &  In-$s$  &  36\\
Cs$_2$KBiCl$_6$  &  11.498  &-&   4.32 &-& Direct  &   Bi-$s$  &  Bi-$p$  & 38 \\
Cs$_2$NaBiBr$_6$ &	11.615	& 11.357 \cite{NaBiBr} &	3.29 & 3.10 \cite{NaBiBr}&	Indirect &	Bi-$s$ &	Bi-$p$ &	38\\
\hline
\end{tabular}
\label{T1}
\end{table*}
As the next step, we perform first-principles calculations to estimate the electronic properties of a set of HDPs. For all HDPs discussed here, we have used A = Cs and X = Cl as an example, and we believe that results and description will not change by replacing A and X site atoms with their equivalent unless the crystal symmetry is destroyed. Since not all HDPs are experimentally synthesized in their pristine phase, we have conceived the hypothetical cubic Fm$\bar{3}$m structure and performed the full relaxation (both atomic positions and crystal lattice parameters) for all of them. The relaxed lattice constant in comparison to the available experimental lattice constant and bandgap calculated with GGA+mBJ+SOC functional is listed in Table \ref{T1}. The obtained band structures (calculated along the path shown in Fig. \ref{Fig2} (b)) are further used to map the B-MOP as shown in Figs. \ref{Fig3} and \ref{Fig4}. Our aim in studying the pristine phases of HDPs is to understand the properties of the end member crystals, which help to predict the properties of the cation mixed phases. 
\section{Construction of Band Projected Molecular-Orbit Picture (B-MOP)}
A MOP examines the possible chemical bondings and provides us with a broader picture of the electronic structure of a material and its universality in a given family. Therefore, without carrying out comprehensive electronic structure calculations, it is possible through MOP to develop an insight into how the modulations in the electronic structure across a family due to chemical substitution and doping. Here, we construct the MOP in three steps: First, the free atomic energy levels of the valence orbitals are estimated using the Hartree-Fock theory, and their relative positions in the energy scale are obtained. In the second step, orbital projected band structures are carried out using DFT. In the third step, the band centers of the projected bands are linked to the free atomic orbital energy levels so as to obtain probable chemical bonding and their strengths and finally draw the MOP.  Such an attempt of linking the schematical MOP with eigenstates and eigenvalues in the momentum space has never been done before. Therefore, to distinguish it from the conventional not-to-scale MOPs, we name it as band projected MOP (B-MOP). \par
\begin{figure*}
\centering
\includegraphics[scale=0.43]{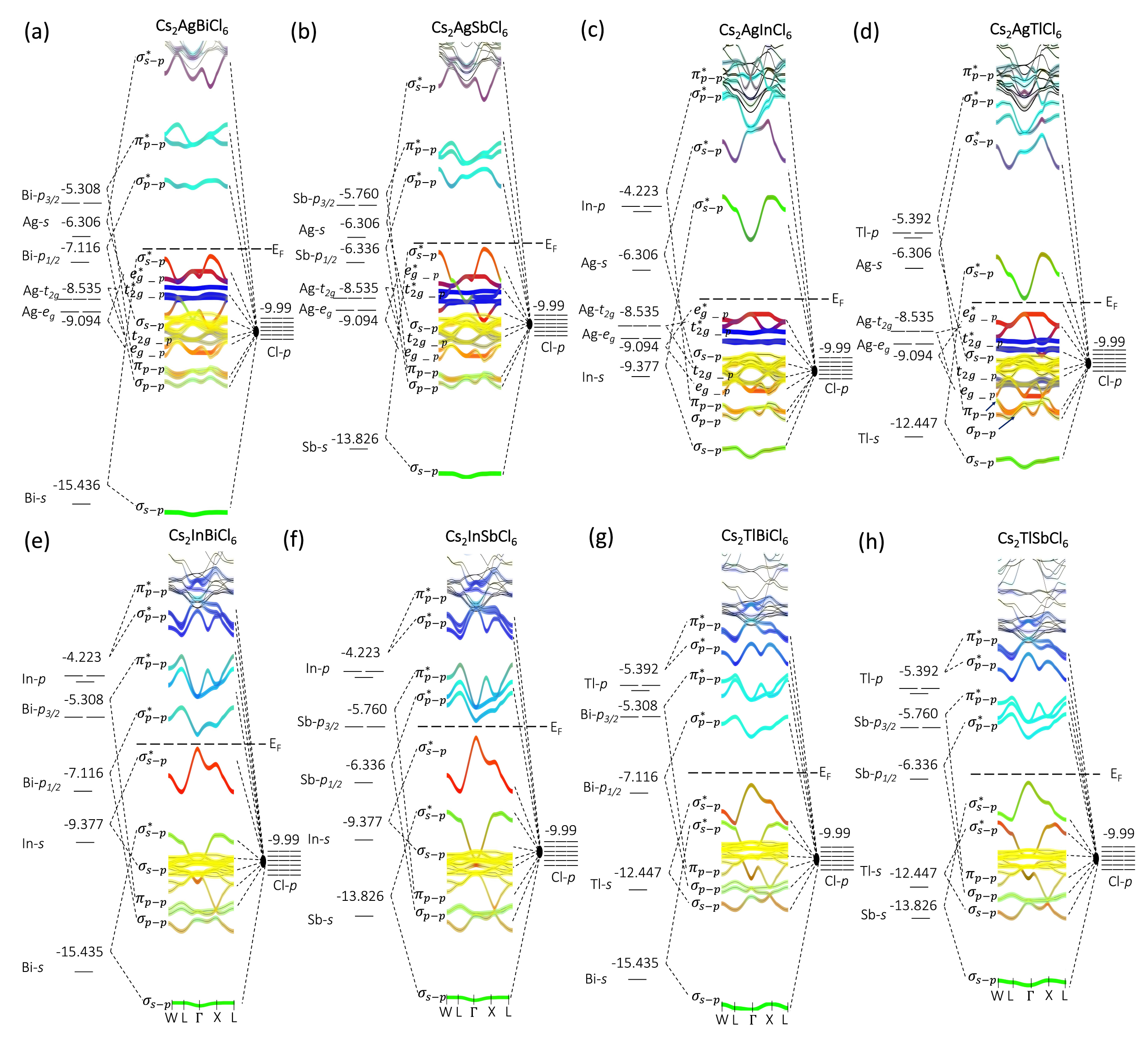}
\caption{The band projected molecular orbital picture (B-MOP) of halide double perovskites as envisaged from the following molecular hybridizations: B(Ag)-\{$s$, $d$\}--Cl-$p$, B(In, Tl)-\{$s$, $p$\}--Cl-$p$ and B$^\prime$(Bi, Sb, In, Tl)-\{$s$, $p$\}--Cl-$p$ atomic orbitals produces the bonding and antibonding orbitals along with the nonbonding Cl-$p$ orbitals.  The free atomic energy levels are estimated from Hartree-Fock's theory. The interactions among the B and B$^\prime$ states are not represented explicitly in these MOPs.}
\label{Fig3}
\end{figure*}

Figure \ref{Fig3} shows B-MOPs for Ag-based HDPs. It infers the formation of bonding and antibonding spectrum arising from \{Ag-($s$, $d$); Bi/Sb/In/Tl-($s$, $p$)\} -- X-$p$ covalent hybridizations. The bonding spectrum consists of $e_g-p$, $t_{2g}-p$, $\sigma_{s-p}$, $\sigma_{p-p}$, $\pi_{p-p}$ interactions, and the antibonding spectrum consists of $e_g-p^*$, $t_{2g}-p^*$, $\sigma_{s-p}^*$, $\sigma_{p-p}^*$, $\pi_{p-p}^*$ interactions. The conservation of basis leaves behind eight non-bonding states combinedly formed by the X-$p$ orbitals. The strength of covalent hybridization is measured by the energy difference between the two corresponding bonding and antibonding pair. The B-MOPs suggest the strength of hybridization in the increasing order as $s-p$ $>$ $p-p$ $>$ $d-p$. Now the valence and conduction band edge states (VBE and CBE) will be determined from the valence electron count (VEC), which is defined as the sum of the valence electrons in the constituent member element. In the case of Cs$_2$AgBiCl$_6$ (Cs$_2$AgSbCl$_6$), the VEC is found to be 48 (see Table \ref{T1} for other HDPs), and therefore the electron filling implies that VBE is formed by Ag-$e_g$ -- X-$p$ hybridized orbitals and CBE is formed by Bi-$p$ -- X-$p$ (Sb-$p$ -- X-$p$) hybridized orbitals. Similarly, for Cs$_2$AgInCl$_6$ (Cs$_2$AgTlCl$_6$), with VEC  46, the VBE is formed by Ag-$e_g$ -- X-$p$ and CBE is formed by In-$s$ -- X-$p$  (Tl-$s$ -- X-$p$) hybridized orbitals, respectively.\par
From B-MOPs (Fig. \ref{Fig3} (a, b)), we describe some of the main features that play an important role in establishing the electronic structure in HDPs. For the Ag-based compounds, these are:
(I) The $t_{2g}$ and $e_g$ orbitals dominated bands are narrow, indicating weaker interactions with the Cl-$p$ states, whereas B/B$^\prime$-$s$ based orbitals dominated bands are wider due to their stronger interactions with Cl-$p$ states. 
(II) Depending on the VEC, Ag-HDPs demonstrate two kinds of bandgaps. For Cs$_2$AgBiCl$_6$ and Cs$_2$AgSbCl$_6$, the anti-bonding spectrum $\sigma^*_{s-p}$ is occupied to form the VBE with VBM at X while $\sigma^*_{p-p}$ constitute the CBE with CBM at $\Gamma$. For Cs$_2$AgInCl$_6$ and Cs$_2$AgTlCl$_6$, which have two less VEC compared to Cs$_2$AgBiCl$_6$ and Cs$_2$AgSbCl$_6$, the $\sigma^*_{s-p}$ is unoccupied to form the CBE with CBM at $\Gamma$  and $e^*_{g-p}$ is occupied to from the VBE with VBM lies on the $\Gamma-X$ flat band.
(III) Bandgap variation: The free atomic energies of B and B$^\prime$-atoms are turning out to be the deterministic factor for the magnitude of the  bandgap as they influence the position of the bands. For example, the bandgap difference of  $\sim$2.64 eV between Cs$_2$AgInCl$_6$ and Cs$_2$AgTlCl$_6$ can be attributed to the free atomic energy difference between In-$s$ and Tl-s orbitals (see Fig. \ref{Fig3} (c, d)). Since the VBM is formed by  $e^*_{g-p}$ bonding for both cases, the bandgap is determined by the position of the $\sigma^*_{s-p}$ led CBM. With In-$s$ energy ~3 eV higher than that of Tl-$s$, the latter forms a narrow bandgap of 0.64 eV, and the former forms a wide bandgap of 3.3 eV.
In the case of Cs$_2$AgBiCl$_6$ and Cs$_2$AgSbCl$_6$, where Bi/Sb-Cl $\sigma^*_{s-p}$ and $\sigma^*_{p-p}$ form the VBE and CBE respectively, the bandgaps are of a nearly similar order. This is due to the fact that both Bi-$s$ and Bi-$p$ energies are lower by a similar magnitude with respect to that of Sb-$s$ and Sb-$p$ energies. Hence, the relative positioning of VBM and CBM are similar for both compounds.\par
Similarly, the salient features that we obtain from B-MOP for the HDPs (Cs$_2$InBiCl$_6$, Cs$_2$InSbCl$_6$, Cs$_2$TiBiBr$_6$, Cs$_2$TlSbBr$_6$ without Ag) are as follows (Fig. \ref{Fig3} (e-h)). (I) All are direct bandgap systems with B-Cl- $\sigma^*_{s-p}$ and B$^\prime$-Cl-$\sigma^*_{p-p}$ forming the VBE and CBE, respectively. (II) When B is In, the system exhibits a narrower bandgap; when it is Tl, it exhibits a wider bandgap. It is largely  attributed to the fact that In-$s$ free atomic energy levels are higher than that of Tl-$s$ by ~3.0 eV. Hence, in the case of the latter, the VBE goes up to increase the separation between CBE and VBE. Furthermore, if we compare the case of B$^\prime$ as Bi and Sb, the former show a smaller bandgap as Bi-$p_{1/2}$ free atomic energy level is lower than that of Sb-$p_{1/2}$ to lower the position of CBE. \par
\begin{figure}
\centering
\includegraphics[angle=-0.0,scale=0.24]{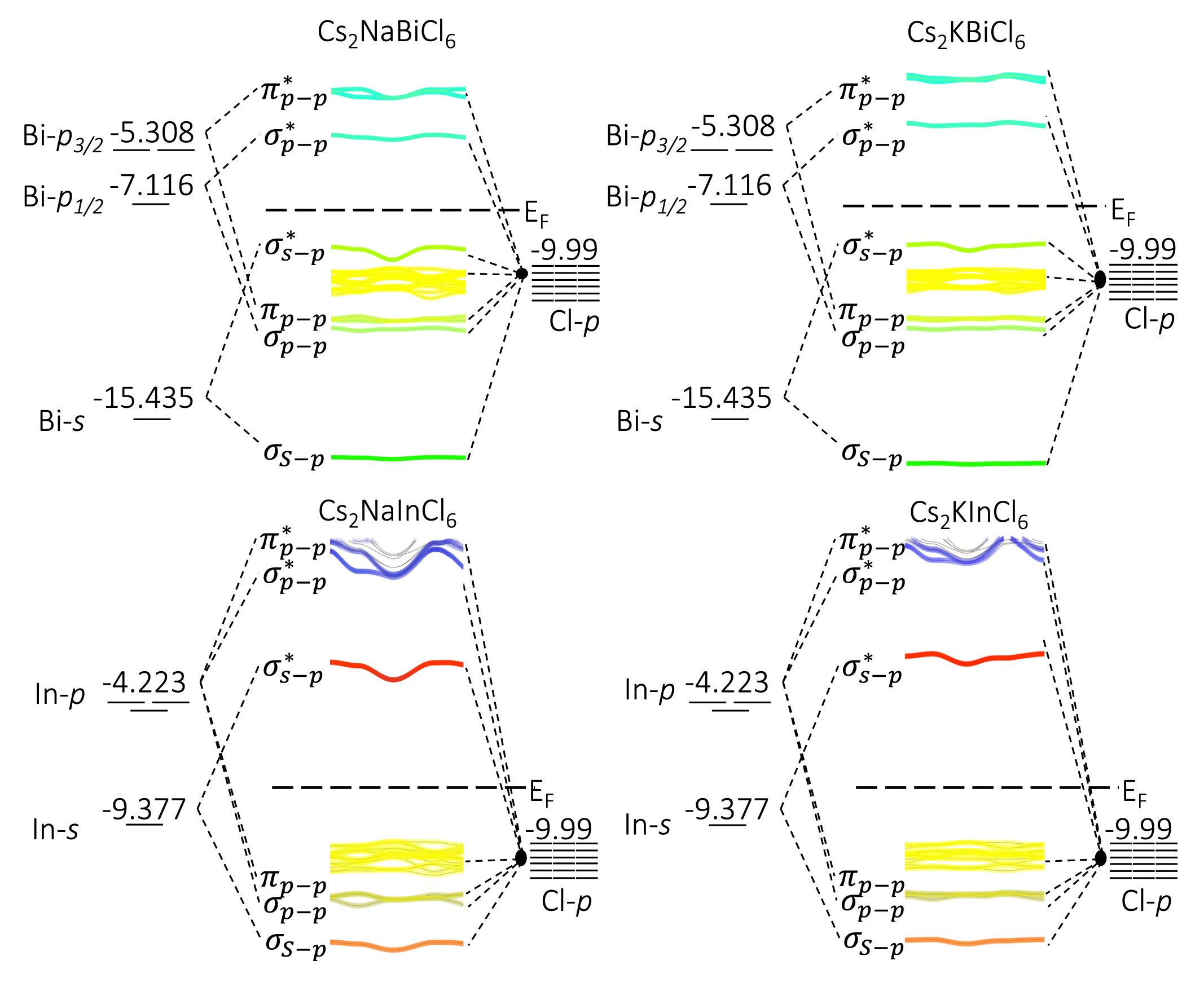}
\caption{The B-MOP of HDPs as envisaged from the molecular hybridization of (Na, K)-\{$s$, $p$\}--X-$p$ and (In, Bi)-\{$s$, $p$\}--X-$p$ atomic orbitals produces the bonding and antibonding orbitals along with the nonbonding X-$p$ orbitals.}
\label{Fig4}
\end{figure}

When the group-1A atoms (Na, K, etc.) occupy the B-site of HDPs, the B-MOP for such compounds is illustrated in Fig. \ref{Fig4}. The B site acts as an electron donor and does not participate in the band formation. The interaction of four B$^\prime$-\{$s$, $p$\} orbitals with 18 Cl-$p$ orbitals give rise to four bonding states ($\sigma_{s-p}, \sigma_{p-p}$, and $\pi_{p-p}$), four corresponding anti-bonding states ($\sigma_{s-p}^*, \sigma_{p-p}^*$, and $\pi_{p-p}^*$) and fourteen flat bands (shown in yellow). When B$^\prime$ is Bi, the VEC is 38, and therefore the anti-bonding bands $\sigma_{s-p}^*$ and $\sigma_{p-p}^*$ form the VBE and CBE respectively. It results in a wide and indirect bandgap system (4.0 - 4.5 eV). There is a narrow separation between $\sigma_{s-p}^*$ and flat bands. When B$^\prime$ is In, the VEC is reduced by two, and the $\sigma_{s-p}^*$ forms the CBE while the Cl-$p$ flat bands form the VBE. These systems exhibit a wide bandgap to the tune of 5.5 - 6.0 eV approximately. B-MOP for four more compounds, namely Cs$_2$AgAsCl$_6$, Cs$_2$CuBiCl$_6$,  Cs$_2$AgGaCl$_6$,  and Cs$_2$CuInCl$_6$ have been studied. The first two compounds  belong to the group of Cs$_2A$gBiCl$_6$ and Cs$_2$AgSbCl$_6$, and the remaining two belong to the group of Cs$_2$AgInCl$_6$ and Cs$_2$AgTlCl$_6$. The detailed analysis of these four compounds is provided in Section XV of SI.\par

The replacement of Cl by Br nearly replicates the B-MOP \cite{Athrey4}. However, since Br-$p$ free atomic energy level is $\sim$ 0.5 eV higher than that of Cl-$p$, we notice a couple of changes in the electronic structure, and the most important of them is the reduction of the bandgap. The cause of it can be explained by taking the example of Cs$_2$AgBiBr$_6$ and Cs$_2$AgBiCl$_6$.
Here, the prominent interactions that define the VBE and CBE are Ag-$d$ -- Br/Cl-$p$ and Bi-$p$ -- Br/Cl-$p$, respectively. We find that the Br-$p$ energy levels are comparable to that of Ag-$d$ energy levels while the Cl-$p$ energy levels lie $\sim$ 0.5 eV below. Therefore, the Ag-$d$ — Br-$p$ interaction is stronger to push the antibonding $e_g^*$ band above for the Br compound as compared to that of the Cl compound. Since there is a large onsite energy difference between Bi-$p$ and Br/Cl-$p$ orbitals ($\epsilon_{(Bi-p)}-\epsilon_{(Br/Cl-p)}$), the positioning of the CBE ($\sigma_{(p-p)}^*$) is less affected by ($\epsilon_{(BrCl-p)}$). Therefore, the CBE does not see a similar shift as the VBE does, and hence, the Br-based HDPs generally see a lower bandgap. Further discussions are made in the supplementary information (SI) (Section XIX).

\section{Band gap improvement using HSE06+G$_0$W$_0$ analysis}
DFT is considered as a reliable tool for calculating the fundamental properties of materials in their ground states.  However, the common exchange-correlation approximations tend to underestimate quasiparticle (QP) bandgaps. To obtain accurate QP energies, Green’s function-based ab-initio many-body perturbation theory (MBPT) can be employed. In the framework of MBPT, the QP eigensystem is determined through the solution of one-particle equations, incorporating a non-local, energy-dependent self-energy operator referred to as $\Sigma$. In practice, $\Sigma$ is frequently approximated as iGW, where W denotes the screened Coulomb interaction and G represents the one-particle Green’s function. Instead of iterative evaluations of $\Sigma$ at each step, G$_0$W$_0$, a computationally more efficient one-shot approach, is frequently employed. In the context of modeling halide perovskites, it has become evident that achieving an accurate representation of the electronic structure hinges on the calculation of many-body QP energies. This approach effectively corrects localized electronic states, reducing the mixing of orbitals between B-e$_g$ or B$^\prime$-s and X-$p$ near the valence band maximum (VBM) and conduction band minimum (CBM). Consequently, the method enhances accuracy by precisely pinpointing the positions of the VBM and CBM, ultimately leading to more reliable bandgap values.

\textit{Methodology:} In the case of HDPs, G$_0$W$_0$ calculations have been performed on top of the orbitals obtained from hybrid exchange functional (HSE06@G$_0$W$_0$) with 0.25 mixing in Hartree-Fock exchange which is included in the VASP \cite{vasp}. In addition to the SOC parameter, we have taken the number of virtual bands to be almost three times the number of occupied bands. The HSE06@G$_0$W$_0$ band spectra are obtained using VASP interfaced with Wannier90 software \cite{MOSTOFI2008685}.

\begin{figure}
    \centering
    \includegraphics[scale=0.24]{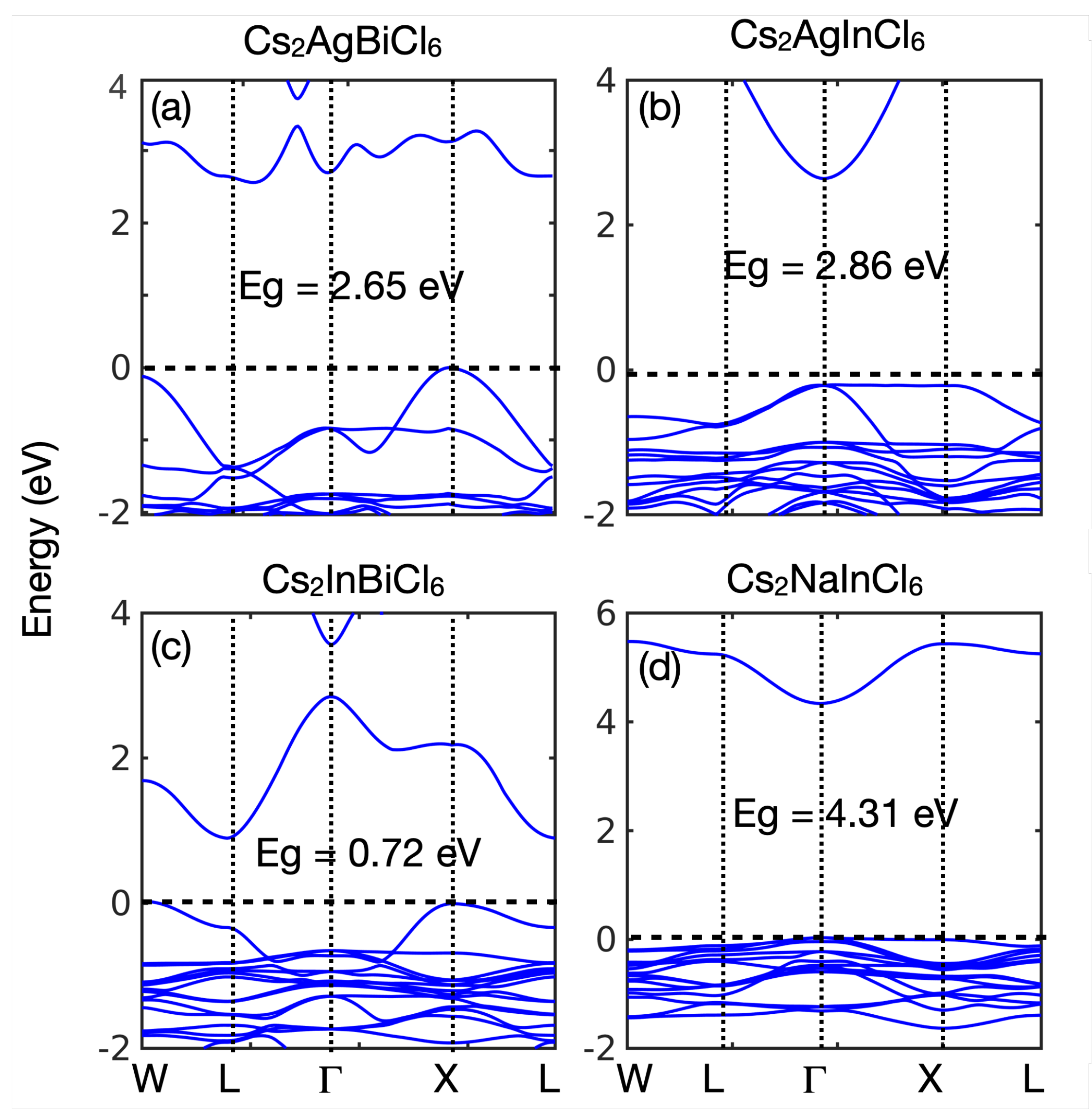}
    \caption{DFT band structure of Cs$_2$AgBiCl$_6$, Cs$_2$AgInCl$_6$,  Cs$_2$InBiCl$_6$, and Cs$_2$NaInCl$_6$ calculated from HSE06@G$_0$W$_0$+SOC method.}
    \label{GW}
\end{figure}

\textit{QP band gap correction:}
While it is computationally expensive to carry out HSE06@G$_0$W$_0$ calculations for all the compounds, for demonstration, we have chosen four compounds, namely Cs$_2$AgBiCl$_6$, Cs$_2$AgInCl$_6$, Cs$_2$InBiCl$_6$, and Cs$_2$NaInCl$_6$, one each from the categories listed in Table \ref{T2}.
 The resulting band gap values from the HSE06@G$_0$W$_0$ method, considering SOC effect, have been listed in Table \ref{GW-table}, and they exhibit a strong agreement with the experimental values. The HSE0@G$_0$W$_0$ obtained band structures are shown in Fig. \ref{GW}.
For materials with indirect semiconducting characteristics like Cs$_2$AgBiCl$_6$, HSE06@G$_0$W$_0$ yields an exact band gap of 2.65 eV. For the direct bandgap semiconductors, Cs$_2$AgInCl$_6$, Cs$_2$NaInCl$_6$, and Cs$_2$InBiCl$_6$ the QP direct gap improves with values 2.86 eV, 4.31 eV and 0.72 eV respectively. Despite this shift in the band gap value, the overall character of the band spectra closely resembles that generated by the GGA-mBJ approach, as shown in Fig. \ref{GW}.
\begin{table}[h]
    \centering
    \caption{Band gap comparison in HDPs calculated using different exchange-correlation functionals. }
    \begin{tabular}{|c|c|c|c|}
    \hline
    Structure&Exp.&GGA-mBJ&HSE@G$_0$W$_0$ \\
    & (eV) &  (eV) & (with SOC) (eV) \\
    \hline
    \hline
         Cs$_2$AgBiCl$_6$&2.65 \cite{athrey1} &2.98 & 2.65\\
         Cs$_2$AgInCl$_6$& 3.02 \cite{AgInCl_gap}& 3.28& 2.86\\
         Cs$_2$InBiCl$_6$ & -&0.53 & 0.72\\
         Cs$_2$NaInCl$_6$ &4.15 \cite{NaInCl-gap} &5.37 & 4.31\\
         \hline
    \end{tabular}
    \label{GW-table}
\end{table}

In Figs. \ref{Fig3} and \ref{Fig4}, we have constructed the B-MOPs by considering the band structures obtained from DFT-GGA+mBJ calculations. Similar B-MOPs can be constructed for the DFT-HSE06@G$_0$W$_0$ calculations by adding a constant energy shift to the conduction band spectrum. This is due to the fact that the band dispersion does not change with HSE06@G$_0$W$_0$.
\section{Construction of Tight-Binding Model Hamiltonian}
While the B/B$^\prime$-X interactions primarily build the MOP and hence provide a broad overview of electronic structure, the next-neighbor B$^\prime$-B interactions influence the band dispersion considerably. As we will see later, the optical properties are driven by these second-neighbor interactions. To provide a deeper insight to the band dispersion, in the following section, we develop a tight-binding (TB) model Hamiltonian involving B-B$^\prime$, B$^\prime$-B$^\prime$, and B-B  interactions. The model is constructed based on linear combination of atomic orbitals (LCAO) and within the framework of Slater-Koster (SK) formalism. In the second quantization notation, the spin-orbit coupled SK-TB Hamiltonian is given as:
\begin{equation}
H = \sum_{i, \alpha} \epsilon_{i\alpha} c^\dagger_{i\alpha}c_{i\alpha} + \sum_{{\braket{ij}};\alpha,\beta} t_{i\alpha j\beta}(c^\dagger_{i\alpha}c_{j\beta}+h.c.)+\lambda \vec{L}\cdot\vec{S},
\end{equation}
where $i$, $j$ and $\alpha$, $\beta$ are indices for the sites and orbitals, respectively. The first term represents the onsite energy ($\epsilon$), while the second term is for the hopping integrals, with $t$ being the hopping strength. The effective tight-binding Hamiltonian matrix includes second nearest-neighbor metal-metal (B-B$^\prime$) interactions, as well as fourth-neighbor B-B and B$^\prime$-B$^\prime$ interactions. The third term in the Hamiltonian represents the atomic spin-orbit coupling (SOC) effect, with the strength of the SOC $\lambda$. \par
\begin{figure*}
\centering
\includegraphics[angle=-0.0,scale=0.2]{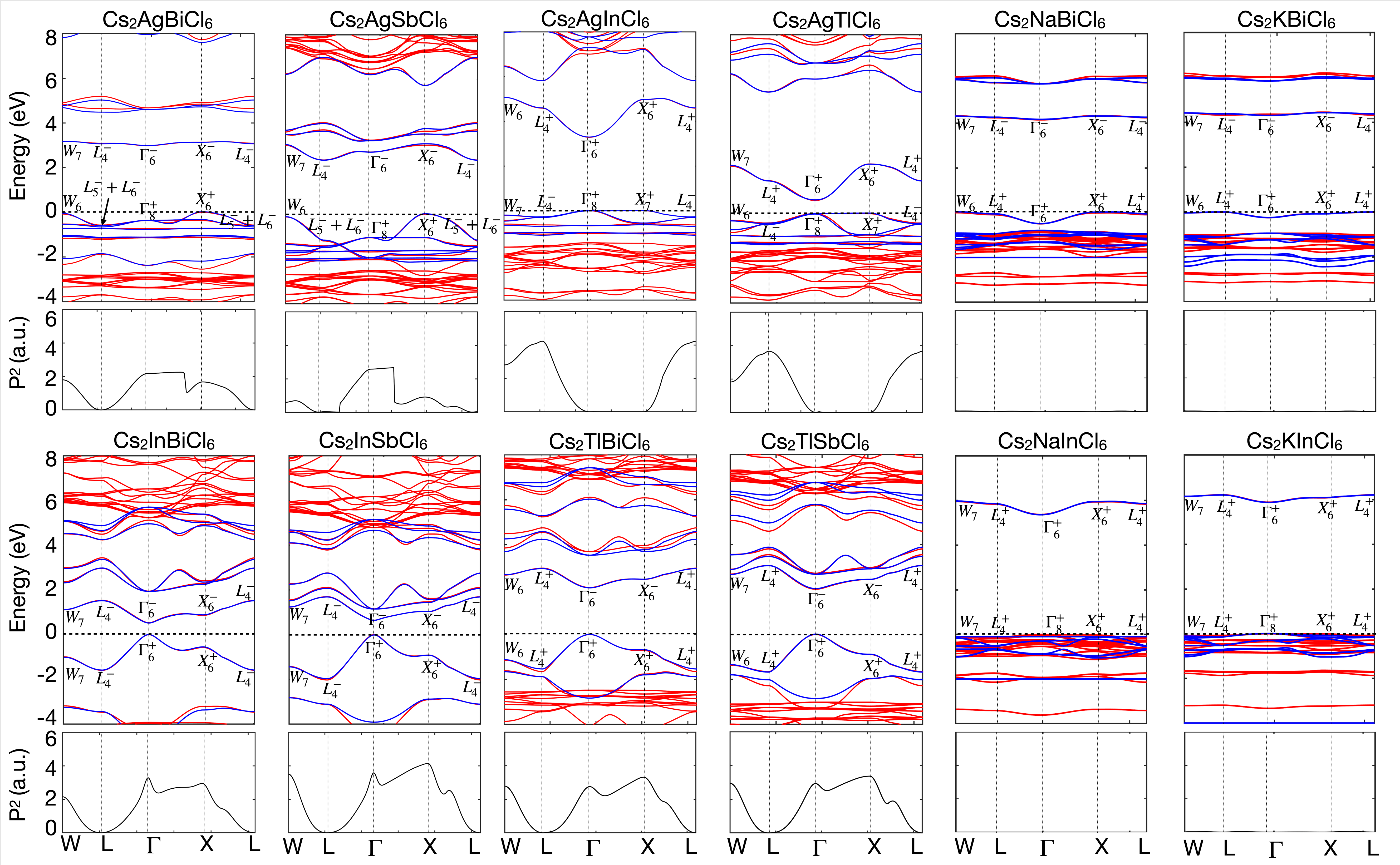}
\caption{The DFT (red) and tight-binding bands (blue) for the HDPs. The squared momentum matrix elements (P$^2$) corresponding to the valence band edge (VBE) to conduction band edge (CBE) transition are also shown for each of the compounds. The parity of the VBE and CBE are mentioned using Koster notations.}
\label{Fig5}
\end{figure*}
A full Hamiltonian involves the interactions among both B and B$^\prime$ valence orbitals as well as X-$p$ orbitals. The present model Hamiltonian with minimal basis is constructed by excluding the X-X interactions while the B-X and B$^\prime$-X interactions are mapped into effective B-B$^\prime$ interactions \cite{Ravi-JCP}. Such a mapping can be validated through Lo\"wdin downfold technique \cite{Lowdin,mayank-2022}. The choice of the basis set is crucial and depends on the atomic orbitals contributing to the bands near the Fermi level. For example, in Ag-based double perovskites, the bands near the Fermi level are contributed from B(Ag)-\{$s$, $d$\} and B$^\prime$-\{$s$, $p$\} orbitals, while in Cs$_2$(In, Tl)(Bi, Sb)Cl$_6$, B-\{$s$, $p$\} and B$^\prime$-\{$s$, $p$\} orbitals contribute to the bands near the Fermi level. The size of the Hamiltonian matrix is dependent on the chosen basis set. The Hamiltonian matrix can be expressed as 
\begin{equation}
    H = \left[\begin{array}{cc}
    H_{BB} & H_{BB'}\\
    H_{BB'}^\dagger&H_{B'B'}\\
    \end{array}\right]
\end{equation}
Here, $H_{BB}$ and $H_{B'B'}$ are the interaction submatrices between the same cations $i.e.$, B-B and B$^\prime$-B$^\prime$ which correspond to the 4$^{th}$ neighbor interactions and $H^{BB'}$/$H^{B'B}$ are interaction matrices between two different cations B-B$^\prime$ and B$^\prime$-B which correspond to the 2$^{nd}$ neighbor interactions (among $d$ and $p$ orbitals for Ag-based HDPs and $p$ - $p$ orbitals interactions for Cs$_2$(In/Tl)(Bi/Sb)Cl$_6$) as shown in Fig. \ref{Fig2} (a) and Fig. S1 in SI. By considering the SOC effect, the basis size is doubled, and sub-matrices take shape:
\begin{equation}
    H_{BB/B'B'} = \left[\begin{array}{cc}
    H^{\uparrow\uparrow} & H^{\uparrow\downarrow}\\
    H^{\dagger\downarrow\uparrow}&H^{\downarrow\downarrow}\\
    \end{array}\right], 
        H_{BB^\prime/B^\prime B} = \left[\begin{array}{cc}
    H^{\uparrow\uparrow} & 0\\
    0 &H_{\downarrow\downarrow}.\\
    \end{array}\right] 
\end{equation}
Here, $H^{\uparrow\uparrow}_{BB/B^\prime B^\prime}$ and $H^{\downarrow\downarrow}_{BB/B^\prime B^\prime}$ are the Hamiltonian sub-matrices corresponding to up and  down spin components and are connected through the time-reversal symmetry. The non-vanishing $H^{\uparrow\downarrow}$ and $H^{\downarrow\uparrow}$ elements of $H_{BB/B^\prime B^\prime}$  are due to the SOC effect.\par
The components of the Hamiltonian matrix, describing the interaction between any two atomic orbitals, say $\alpha$ at site position $\vec{R}_i$ and $\beta$ at site at position $\vec{R}_j$, is calculated using SK integrals ($f_{\alpha\beta}$) \cite{Slater} 
\begin{eqnarray}
    h_{\alpha\beta}^{ij}(k) &=& f_{\alpha\beta}(t; l, m, n)e^{i\vec{k}\cdot(\vec{R}_j-\vec{R}_i)}, \\
        h_{\alpha\beta\sigma\sigma^\prime}(k) &=& \sum_{\braket{j}} h_{\alpha\beta}^{ij}(k)\delta_{\sigma\sigma^\prime}, \\
        H_{\alpha\beta\sigma\sigma^\prime} &=& h_{\alpha\beta\sigma\sigma^\prime} + h_{\alpha\beta\sigma\sigma^\prime}^{SOC}
\end{eqnarray}
The $f_{\alpha\beta}(t; l, m, n)$ depend on the tight-binding hopping parameters $t$ and direction cosines ($l$, $m$, $n$) connecting the site $j$ to $i$. The $h^{SOC}_{\alpha\beta\sigma\sigma^\prime}$ is driven by the atomistic SOC $\lambda\boldsymbol{L\cdot S}$. The required $h_{\alpha\beta\sigma\sigma^\prime}$  as well as $h_{\alpha\beta\sigma\sigma^\prime}^{SOC} $ depend on the basis chosen, which varies from compound to compound, see Section XIII of SI for details.  As shown in Fig. \ref{Fig5}, the bands obtained from the model Hamiltonian are fitted with that of the DFT, and the resulted parameters are listed in Table S10, S11, and S12 of Section XIV of SI. \par
Some of the critical inferences obtained by analyzing the TB parameters are as follows: (I) The 4$^{th}$ nearest-neighbor (B-B/B$^\prime$-B$^\prime$) hopping interactions are very weak (0 - 50 meV) suggesting that the dispersion is mainly driven through the 2$^{nd}$ nearest-neighbor B-B$^\prime$ interactions which are 1-2 order higher in strength. 
(II) Only when Ag and Bi/Sb occupy B and B$^\prime$ sites, the B-$s$ -- B$^\prime$-$s$ interactions become negligible due to large onsite energy differences. In the rest of the members, this interaction is significant enough to influence the dispersion of VBE and CBE (See the B-MOP in Fig. \ref{Fig3}).
(III) For HDPs where the ionic Na$^+$ and K$^+$ occupy the B-site, the dispersions are due to B$^\prime$-B$^\prime$ and B-X interactions. 
(IV) The SOC strength is estimated for both B and B$^\prime$ for each of the compounds considered (see Table S10, S11, and S12 in SI). For Bi and Tl, it is $\sim$0.5 eV, and for other B and B$^\prime$ elements, it is $\sim$0.2 eV. Interestingly, these numbers are comparable to the hopping interactions $t_{\alpha\beta}$. Tight-binding bands for Cs$_2$AgAsCl$_6$, Cs$_2$CuBiCl$_6$,  Cs$_2$AgGaCl$_6$,  and Cs$_2$CuInCl$_6$ are provided in the SI, Section XV. \par
Furthermore, our findings indicate that while the SOC of a compound has a deterministic effect on its bandgap, it does not influence the parity eigenvalues of its VBE and CBE.
\section{Optical Properties Calculation}
The optical absorption coefficient $\alpha(\omega)$ of a material is determined by its frequency ($\omega$) dependent dielectric constant ($\epsilon(\omega) = \epsilon_1(\omega)+i\epsilon_2(\omega)$): 
\begin{equation}
    \alpha (\omega) = \omega \sqrt{\frac{-\epsilon_1(\omega) + \sqrt{\epsilon_1^2(\omega) + \epsilon_2^2(\omega)}}{2}},
\end{equation}
\begin{equation}
\begin{split}
    \epsilon_1(\omega) = 1+\frac{2}{\pi}C \int_0^{\inf}  \frac{\omega^\prime\epsilon_2(\omega^\prime)}{\omega^{\prime 2}-\omega^2}d\omega^\prime, \\\\ \nonumber
    \epsilon_2(\omega) = \frac{e^2\hbar^2}{\pi m_e^2\omega^2}\sum_{v,c} \int_{BZ}d^3k|P_{v,c}|^2\times \\
    \delta(E_c(\vec{k})-E_v(\vec{k})-\hbar\omega).
\end{split}
\end{equation}
Here, C is the Cauchy principal value of the integral; $e$ and $m_e$, respectively, are charge and mass of an electron. P$_{v,c}$ in the expression of $\epsilon_2(\omega)$ are the MME corresponding to a transition from valence band at energy $E_v$ to conduction band at energy $E_c$. The Dirac-delta function switches on the MME contribution when a transition occurs from one state to another. The MME for periodic Block functions $ u_{\vec{k}\beta}$ $ e^{i\vec{k}\cdot \vec{R}}$ are obtained as follows \cite{Lee-2018}:
\begin{equation}
    \bra{\vec{k}, \beta}\vec{P}\ket{\vec{k}, \beta^\prime} = \frac{m_e}{\hbar}\bra{u_{\vec{k}\beta}}\frac{\partial H(\vec{k})}{\partial \vec{k}}\ket{u_{\vec{k}\beta^\prime}}.
\end{equation}
In the case of optical transition, the transitions from the top valence to the bottom conduction band are generally considered. Therefore, the component of relevant MME can be expressed as follows,
\begin{equation}
    (P_{v,c})_{x,y,z} = \frac{m_e}{\hbar}\sum_{\beta,\beta^\prime} u_{\vec{k}\beta,c} \frac{\partial H_{\beta\beta^\prime}}{\partial k_{x,y,z}} u_{\vec{k}\beta^\prime,v}.
\end{equation}
Here, $u_{\vec{k}\beta,c}$ and $u_{\vec{k}\beta^\prime,v}$ respectively represent the eigenvectors associated with the energy eigenvalues $E_v$ and $E_c$. We have calculated the squared MME along the same high symmetric $k$-path as that of the band structure. A detailed derivation to calculate the optical properties of materials using the SK-TB model is given in  Section XVI of the SI.\par
In Fig. \ref{Fig5}, we have shown the band structures for a series of HDPs, and for each of them, the parity eigenvalues (indicated through Koster notations) of VBE and CBE are estimated at the high symmetry points, as shown in the Figure. Corresponding to each of the band structures, the P$^2$ (VBE $\rightarrow$ CBE) is also plotted. According to Laporte’s rule \cite{Laporte:25}, which is applied only for the centrosymmetric systems as is the case here, the optical transition between the VBE and CBE at any given k-point is allowed only when they have opposite parities (odd-even). The validation of it comes through the P$^2$ plot. 
oing beyond the simplistic parity analysis, a thorough group theoretical symmetry analysis is also carried out in Sections XI and XII in the SI to elaborate the selection rules governing the optical transition.\par

 For example, in the case of  Cs$_2$AgBiCl$_6$ and Cs$_2$AgSbCl$_6$, the transition between VBE and CBE at the high symmetry points W, $\Gamma$, and X is allowed due to opposite parities while it is forbidden at L due to same parity (odd-odd). Agreeing with it, the P$^2$ value vanishes at L and is finite elsewhere. In the case of Cs$_2$AgInCl$_6$ and Cs$_2$AgTlCl$_6$, the P$^2$ is zero along the path $\Gamma$-X as both the CBE and VBE have even parity along this path. Therefore, even though these two systems have a direct bandgap, the lowest energy transitions are not allowed, and hence the optical bandgap (defined through transition at L) differs from the electronic bandgap. The compounds Cs$_2$(In/Tl)(Bi/Sb)Cl$_6$ are direct bandgap systems with VBM and CBM lying at $\Gamma$. The finite value of P$^2$ implies the lowest energy allowed transition, which is further verified by the parity analysis. Hence, these compounds are much more promising for optoelectronic applications. In the case of Cs$_2$(Na/K)BiCl$_6$, where the VBE and CBE are formed by Bi-$s$ and Bi-$p$ respectively (see Fig. \ref{Fig4}), while the opposite parity allows the optical transition between VBE and CBE, the P$^2$ is found to be negligible (of the order $\sim$ 10$^{-2}$). On the other hand, the same parities for VBE and CBE of Cs$_2$(Na/K)InCl$_6$ forbid the optical transition, and expectedly, P$^2$ is found to be zero. We may note that, since these systems are ionic with large charge transfer from Na/K to Cl, the transition dipole moment for this VBE to CBE transition is naturally weak \cite{Manna-2019}.  \par
\begin{figure}[ht]
\centering
\includegraphics[scale=0.25]{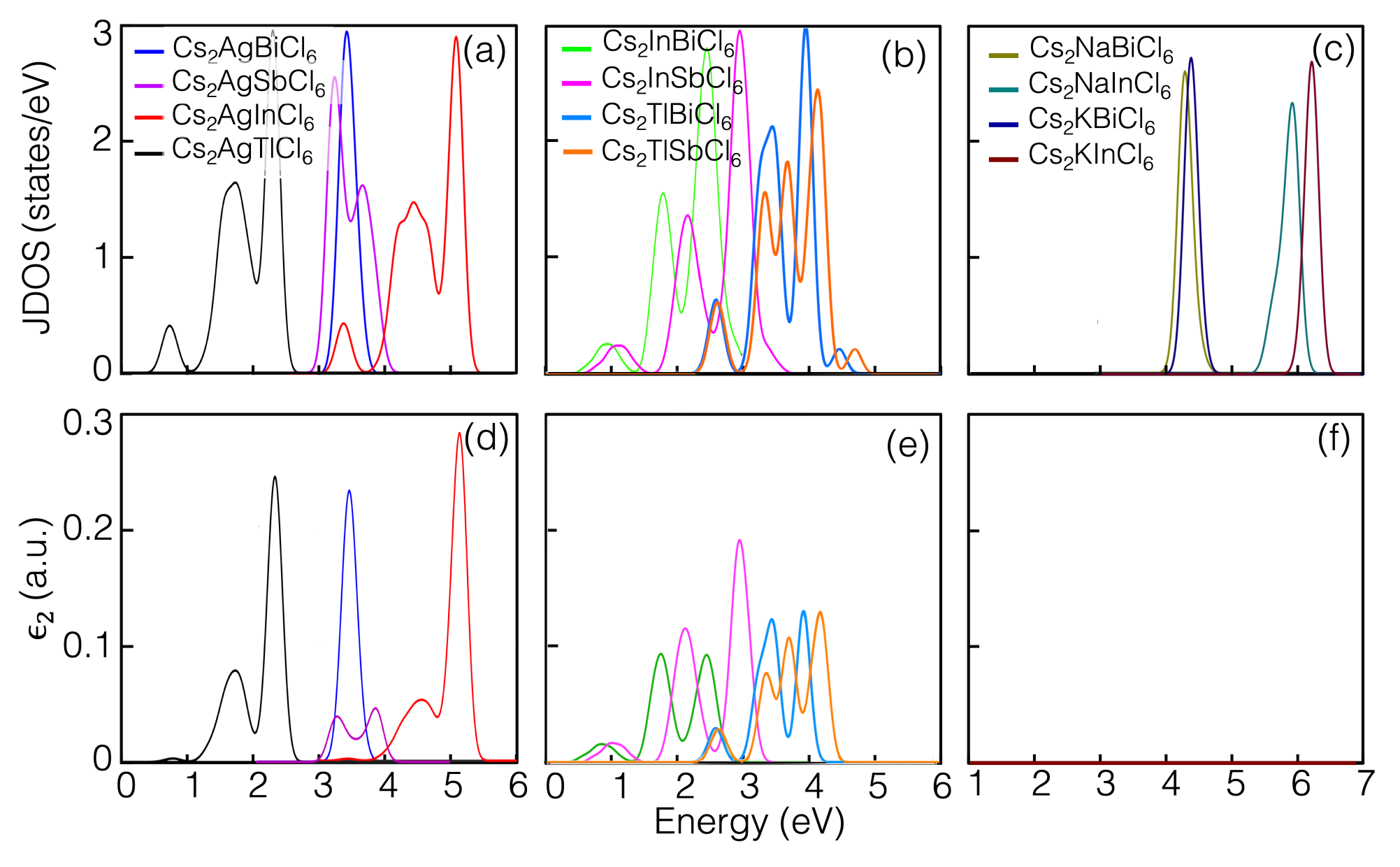}
\caption{(a-c) Model Hamiltonian obtained joint densities of states (JDOS) of HDPs. The JDOS is calculated for the transition between VBE and CBE. (d-f) Calculated $\epsilon_2(\omega)$ showing the effect of parity forbidden transition and effective optical bandgap of these compounds.}
\label{Fig6}
\end{figure}

The optical transition analysis through Fig. \ref{Fig5} is based on a given $k$-path (W-L-$\Gamma$-X-L), which does not necessarily provide the complete picture to understand the transition due to the whole BZ, we have calculated the joint densities of states (JDOS). The JDOS provides a measure of the number of all possible optical transitions between the occupied valence band and the unoccupied conduction band separated by photon energy $\hbar \omega$.
\begin{equation}
        JDOS = \frac{e^2\hbar^2}{\pi m_e^2\omega^2}\sum_{v,c} \int_{BZ}d^3k \hspace{0.1cm}\delta(E_c(\vec{k})-E_v(\vec{k})-\hbar\omega).
\end{equation}
The JDOS, obtained for the lowest electronic transition, is plotted in the upper panel of Fig. \ref{Fig6}. Also, $\epsilon_2 (\omega)$, which can be best described as the MME modulated JDOS, is shown in the lower panel of Fig. \ref{Fig6}. \par
As discussed in the above paragraph, the optical transition in Na/K based compound is negligible (see Fig. \ref{Fig6} (f)) even though the JDOS (Fig. \ref{Fig6} (c)) shows some transition probabilities between 4 to 6.5 eV. This is due to the fact that in these ionic systems, the dipole-dipole transition is very weak. The $\epsilon_2(\omega)$ gives a measure of optical bandgap and how it differs from electronic bandgap (as inferred from the JDOS alone). For the case of Cs$_2$InBiCl$_6$, Cs$_2$InSbCl$_6$, Cs$_2$TlBiCl$_6$, and Cs$_2$TlSbCl$_6$, peaks of JDOS and $\epsilon_2$ suggest the direct and strong optical transitions (see Fig. \ref{Fig6} (b and e)). In Cs$_2$AgTlCl$_6$, the first small peak in JDOS below 1 eV is suppressed in  $\epsilon_2(\omega)$, implying the optical bandgap is $\sim$ 1 eV (see Fig. \ref{Fig6} (a, d)). Because of a similar reason, in the case of Cs$_2$AgInCl$_6$ the optical bandgap is estimated to be $\sim$ 3.6 eV. Cs$_2$AgBiCl$_6$ and Cs$_2$AgSbCl$_6$ have large JDOS values however their $\epsilon_2$ curves show rather different features. The optical transition for Cs$_2$AgSbCl$_6$ is quite weak compared to Cs$_2$AgBiCl$_6$. 
\section{Tailoring of the optoelectronic properties: Doping on cationic sites }
\begin{figure}[h]
\centering
    \includegraphics[angle=-0.0,scale=0.255]{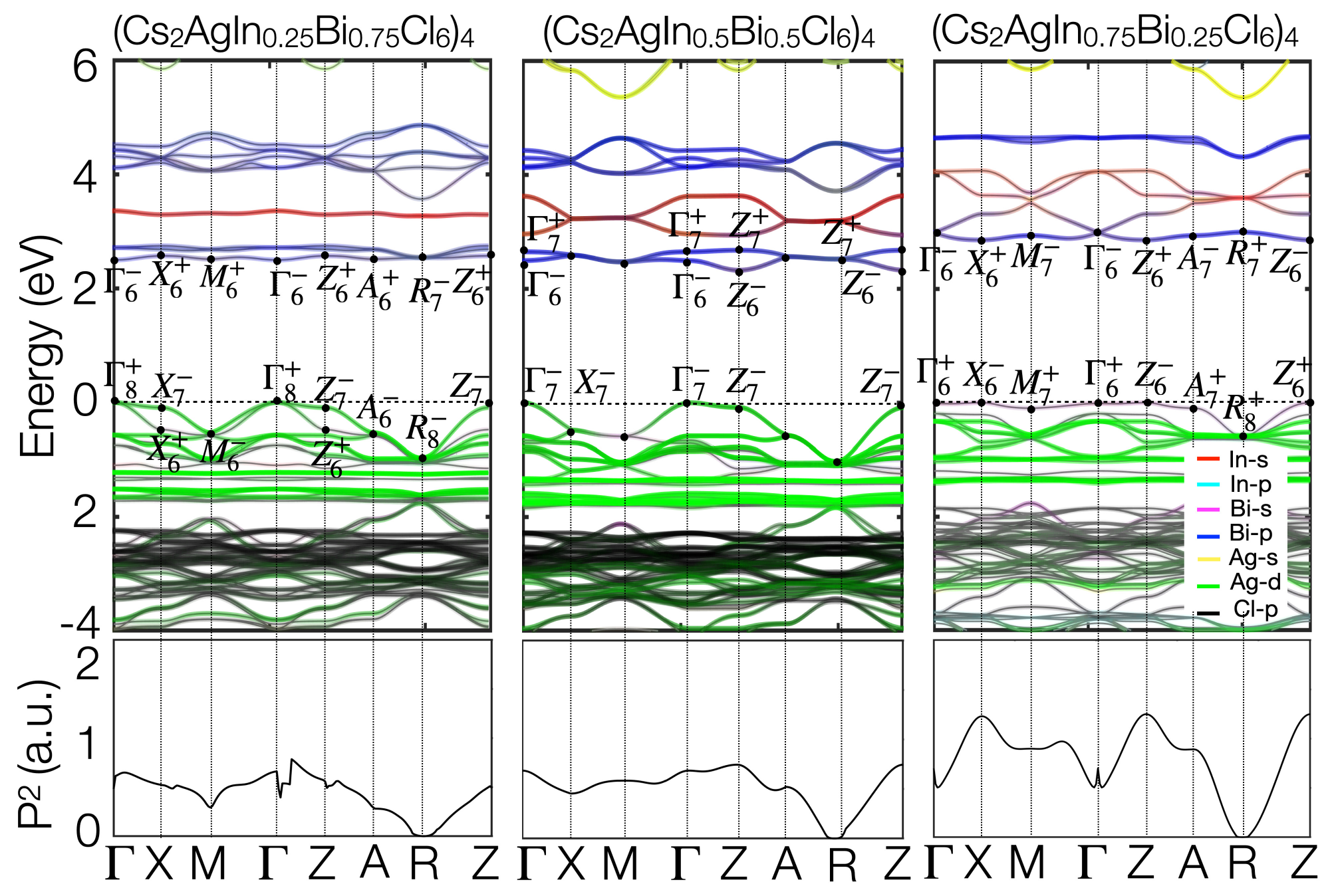}
\caption{(Upper panel) DFT obtained orbital projected band structures of Cs$_2$AgIn$_x$Bi$_{1-x}$Cl$_6$.  The calculations are performed with a structurally relaxed four-formula unit supercell. (Lower panel) The P$^2$ plot, which is obtained from the TB model Hamiltonian designed for the supercell (see SI).}
\label{Fig7}
\end{figure}
\begin{figure}[ht]
\centering
 \includegraphics[angle=-0.0,scale=0.255]{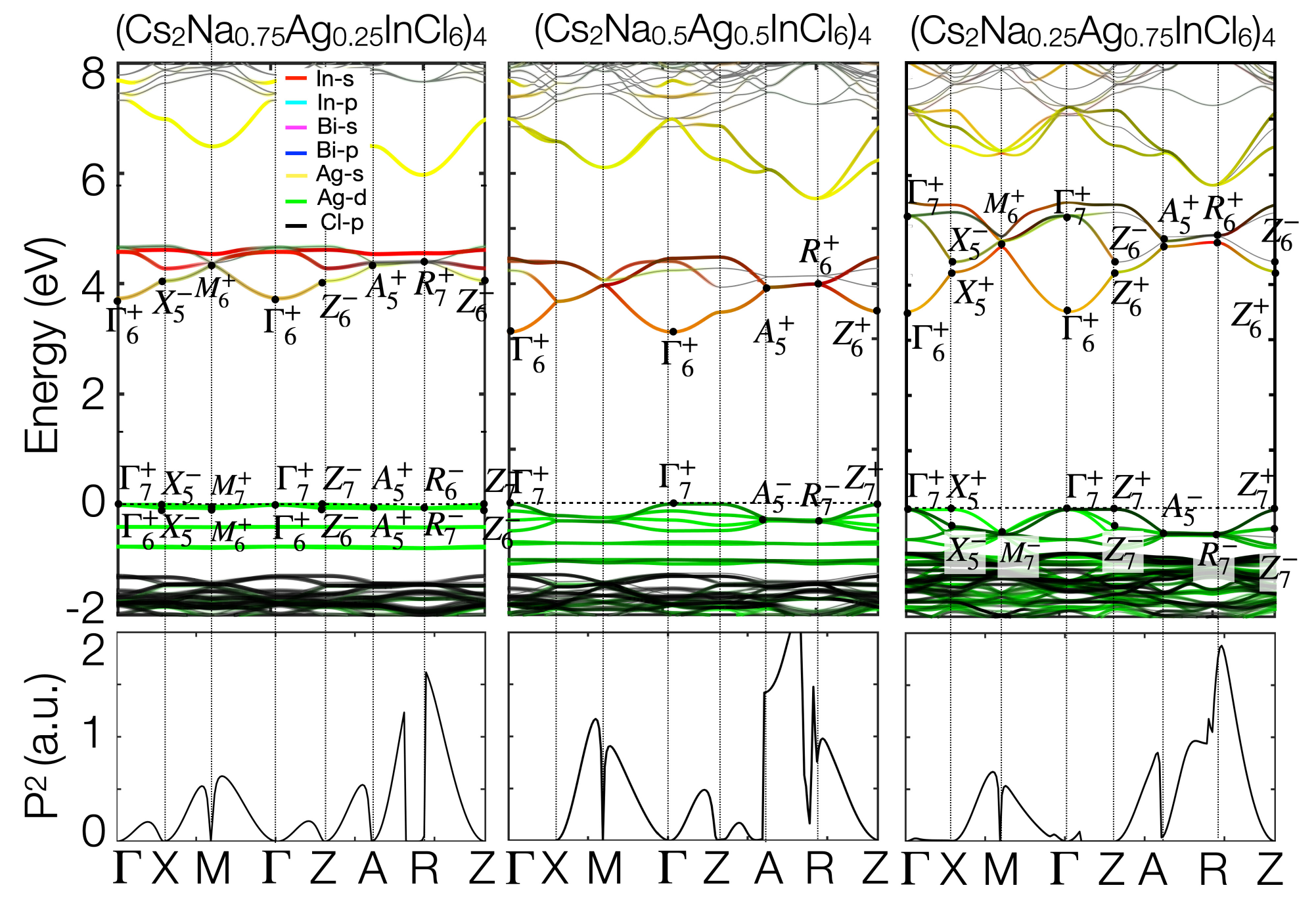}
\caption{Same as Figure \ref{Fig7} but for the compound Cs$_2$Na$_{1-x}$Ag$_x$InCl$_6$.}
\label{Fig8}
\end{figure} 
With a good understanding of the electronic structure and optical behavior of the pristine HDPs, in this section, we will show how the optoelectronic properties can be tailored by cation intermixing. For demonstration, we have considered two cases (a) Cs$_2$AgIn$_x$Bi$_{1-x}$Cl$_6$ and (b) Cs$_2$Na$_{1-x}$Ag$_x$InCl$_6$.  A detailed description of the TB model for cation-intermixed HDPs supercells and corresponding TB bands fitted DFT bands are provided in Sections XVII, XVIII, and Fig. S4 of the SI, respectively. In the first case, the end member Cs$_2$AgBiCl$_6$ has an indirect bandgap (unfavorable for optical transition), and Cs$_2$AgInCl$_6$ has a direct bandgap, however, with parity forbidden transitions. Furthermore, the CBE and VBE of the former are made up of Bi-$p$ and Ag-$e_g$/Bi-$s$ orbitals, respectively, while for the latter, these are made up of In-$s$ and Ag-$e_g$ orbitals, respectively. A careful look at the B-MOP suggests that with Bi and In intermixing, the CBE will be dominated by Bi-$p$ while VBE will be by Ag-e$_g$ states as well as dopant Bi-$s$ states. Upon dilution, the Bi-states are expected to be localized and  alter the shape of the corresponding band. Thereby, the optical absorption spectra are also expected to change. To verify, in Fig. \ref{Fig7}, we have plotted the DFT band structure of  Cs$_2$AgIn$_x$Bi$_{1-x}$Cl$_6$ ($x$ = 0.25, 0.5, and 0.75) and  P$^2$,  which are estimated from the model Hamiltonian. The salient features of the band structures are as follows: (I) With increasing In, the Bi-$s$ state dominates the VBE while the Ag-e$_g$ dominated bands are pushed below. (II) The CBE is always formed by the Bi-$p$ characters. However, its shape changes with doping concentration, which implies new interactions, Bi-$p$ and In-$s$ states. (III) For the case of  $x$ = 0.75, both VBE and CBE are narrower across the BZ path and also have a direct bandgap with the VBM and CBM lying at Z. Interestingly, at every high symmetry point (except at R), VBE and CBE have opposite parities to allow optical transitions across the path. The P$^2$ plot further substantiates it.  \par
Similarly, in the second case, we are doping Ag at the Na site of Cs$_2$NaInCl$_6$. As already been discussed,  the Cs$_2$NaInCl$_6$ does not exhibit any optical transition as the P$^2$ vanishes, thanks to the identical parity of CBE and VBE  (see Fig. \ref{Fig5}). Fig. \ref{Fig8} shows the orbital projected band structure of Cs$_2$Na$_{1-x}$Ag$_x$Cl$_6$ ( $x$ = 0.25, 0.5, and 0.75) and the corresponding P$^2$ plot. The salient features of the band structures are as follows: (I) The Ag-$d$ characters tend to dominate the VBE, and for diluted Ag concentration, the VBE becomes flatter like the impurity bands. (II) The shape of CBE remains largely unchanged though it becomes wider with increasing Ag concentration. (III) The parities of CBE and VBE have altered at certain $k$ points, and therefore, P$^2$ is no longer vanishing. Recent studies by Luo $et. al.$ \cite{Luo2018}, report the existence of self-trapped excitons (STEs) and, thereby, broadband and white light emission in the Cs$_2$Na$_{1-x}$Ag$_x$Cl$_6$.  The present study suggests that the STEs are formed by the flat bands, and the localized carrier at the Ag-site becomes the source of the white light emission. \par
\begin{figure}
    \centering
    \includegraphics[scale=0.33]{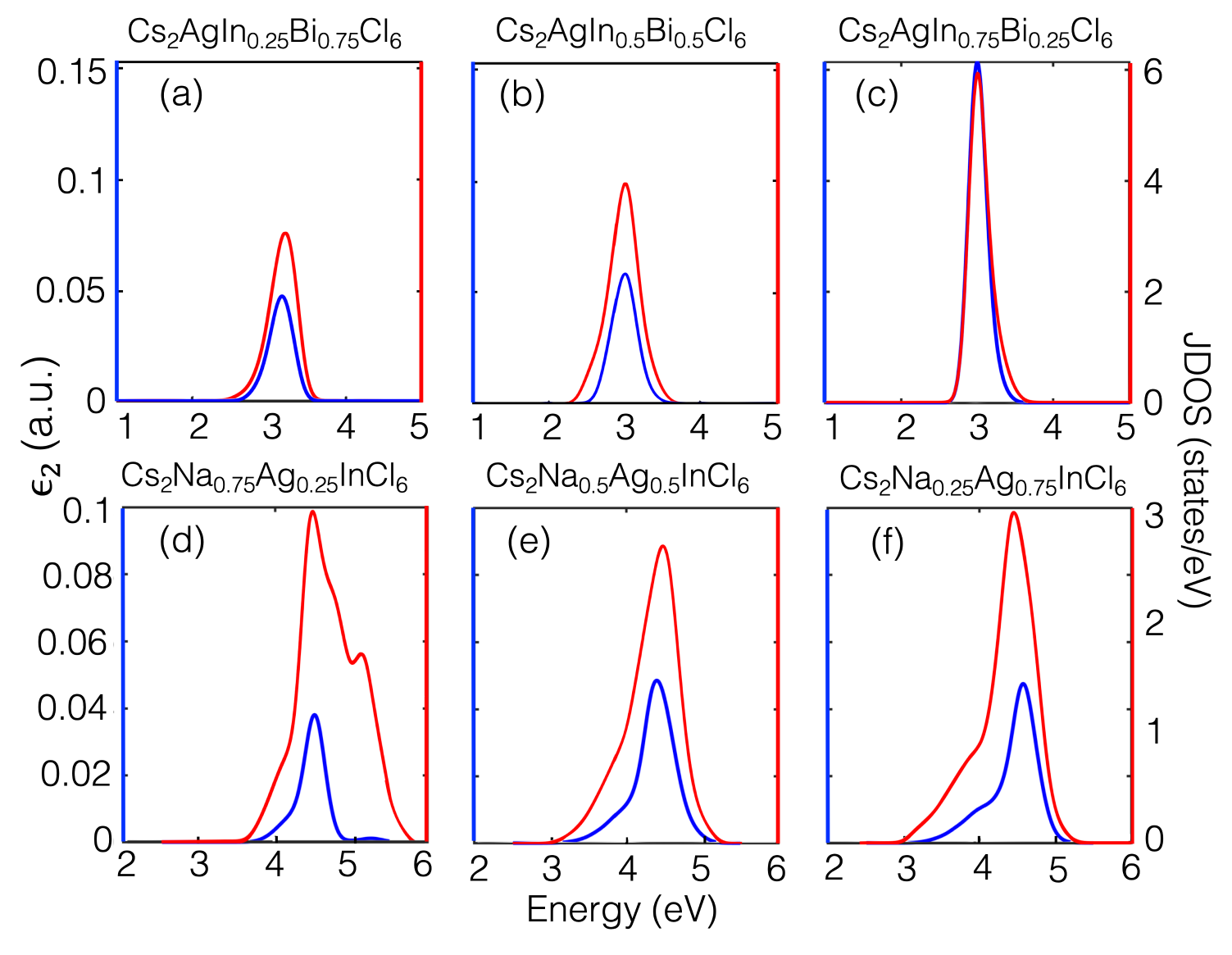}
    \caption{JDOS (shown in red) and imaginary part of the dielectric constant (shown in blue) for cation intermixed Cs$_2$AgIn$_{x}$Bi$_{1-x}$Cl$_6$ (upper panel) and Cs$_2$Na$_{1-x}$Ag$_x$InCl$_6$ (lower panel). }
    \label{Fig9}
\end{figure}
\begin{figure}
\centering
 \includegraphics[angle=-0.0,scale=0.4]{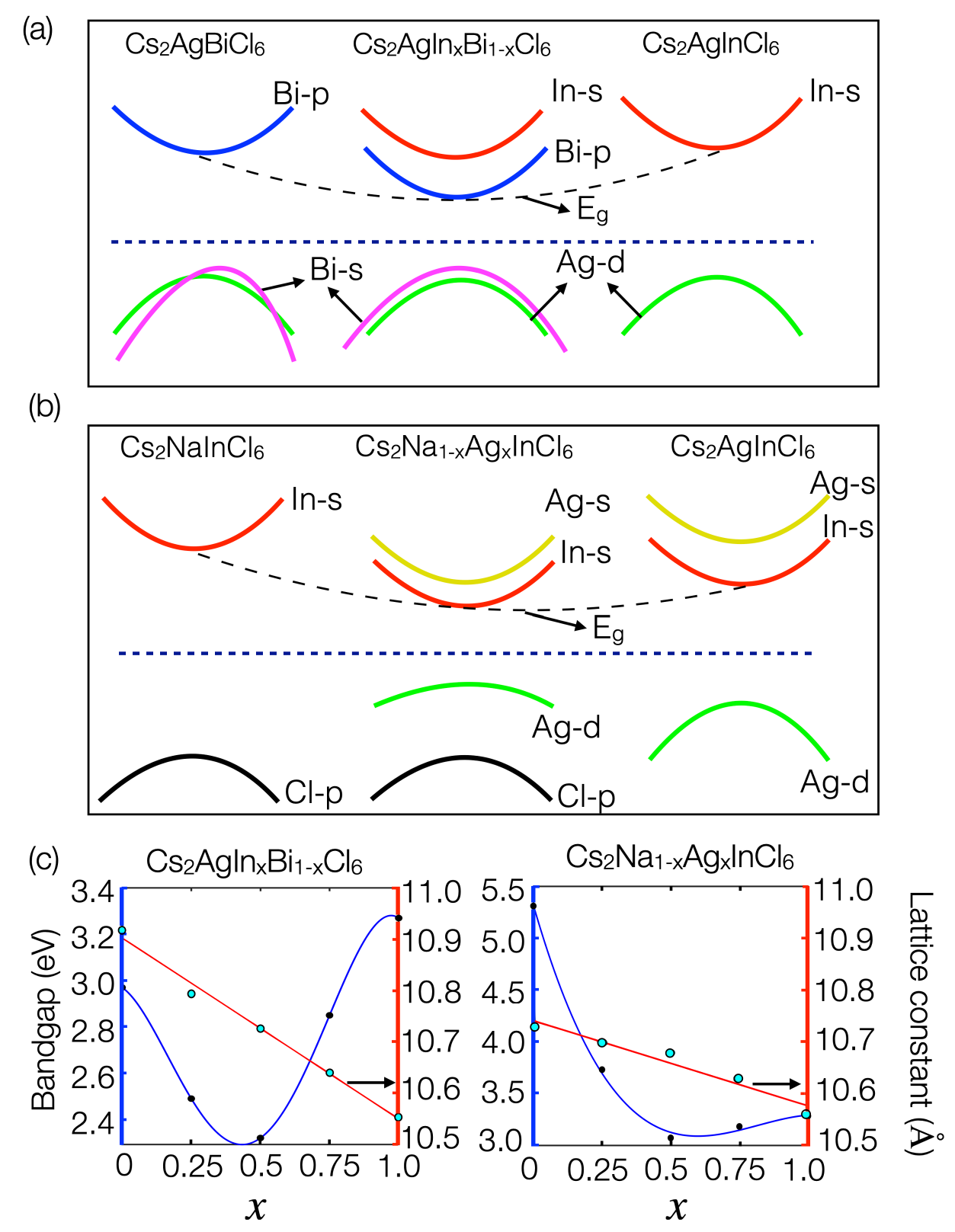}
\caption{(a) The schematic illustration of the orbital resolved band structure in the vicinity of Fermi level and demonstration of bandgap bowing in Cs$_2$AgIn$_{x}$Bi$_{1-x}$Cl$_6$ and Cs$_2$Na$_{1-x}$Ag$_x$InCl$_6$. (b) The bandgap (E$_g$) as a function of doping concentration $x$ in Cs$_2$AgIn$_x$Bi$_{1-x}$Cl$_6$ (left) and Cs$_2$Na$_{1-x}$Ag$_x$InCl$_6$ (right). The DFT obtained bandgap values are shown in black circular dots, and the polynomial fitted curves are shown in blue. The variation of the lattice constant ($a$) as a function of $x$ is also shown. }
\label{Fig10}
\end{figure}
The above analysis on cation intermixed HDPs is made by examining the eigenspectrum along certain k-paths. To substantiate the conclusions made, we now consider the full BZ and estimate JDOS and $\epsilon_2$ for Cs$_2$AgIn$_{1-x}$Bi$_x$Cl$_6$ and Cs$_2$Na$_{1-x}$Ag$_x$InCl$_6$ in Fig. \ref{Fig9}. The red curve represents the JDOS for the transition from VBE to CBE, and the blue curve represents $\epsilon_2$. In the former case, we indeed observed increasing JDOS and $\epsilon_2$ with increasing In concentration.  as shown in Fig. \ref{Fig9} (a-c). The direct and strong optical transition makes the system highly photoluminescent, which agrees with the recent experimental studies by Appadurai $et. al.$ \cite{Appadurai2021}. In the case of  Cs$_2$Na$_{1-x}$Ag$_x$InCl$_6$, any amount of Ag doping induces optical transition.\par
\section{Bandgap bowing effect}
The band positioning due to cation doping, as discussed in the previous section, can be schematically summarized through Figs. \ref{Fig10} (a, c). It shows that with doping, not only there is a reconstitution of VBE and CBE, but also there is a shift of these edge bands. Either the VBE goes up, or CBE comes down, or both happen simultaneously to reduce the bandgap with respect to the parent pristine compounds. This effect, in general, is called bandgap bowing which occurs less often than the linear change in the bandgap as defined by Vegard's law. To quantify the bandgap bowing, in Fig. \ref{Fig10} (c), we have estimated the bandgap as a function of doping concentration in Cs$_2$AgIn$_{x}$Bi$_{1-x}$Cl$_6$ and Cs$_2$Na$_{1-x}$Ag$_x$InCl$_6$. The cause of bandgap bowing in the HDPs is a matter of debate in experimental and density functional studies \cite{athrey1, athrey2, aswani-2019, Bowing-2022, HanDan, Bowing-2022, Goyal-2018}. These studies collectively proposed three probable factors for bandgap bowing: These are (I) change in lattice constant, (II)  octahedral distortion, and (III) chemical effect. Through Figs. \ref{Fig7}, \ref{Fig8} and \ref{Fig10} (a, b), and related discussion, we have already discussed how the chemical effect plays a role in band repositioning. Our B-MOP indeed has shown that the free atomic orbital energies play a major role in determining the position of VBE and CBE.
A similar observation was made by D. Han $et. al.$ \cite{HanDan}.\par
\begin{figure}[h]
    \centering
    \includegraphics[scale=0.4]{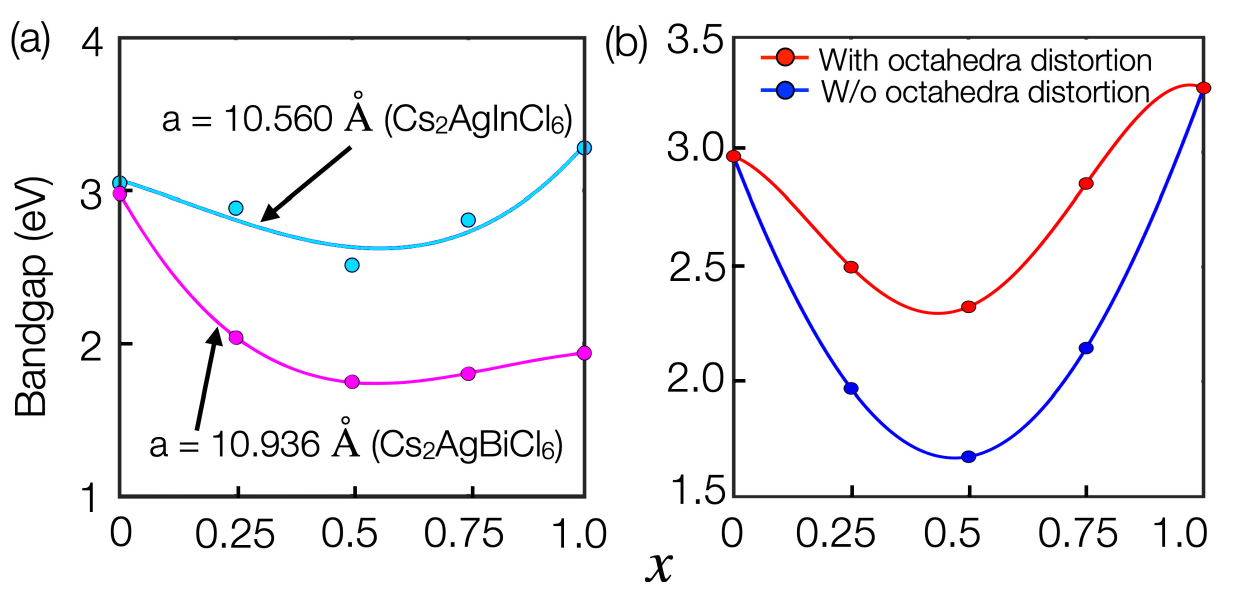}
    \caption{(a) E$_g$ for Cs$_2$AgIn$_{x}$Bi$_{1-x}$Cl$_6$ as a function of $x$, which are estimated by using the lattice parameter of Cs$_2$AgInCl$_6$ (10.56 \AA; solid cyan line) and  Cs$_2$AgBiCl$_6$ (10.936 \AA; solid magenta line) and keeping the structures unrelaxed. (b) E$_g$ as a function of $x$ for the volume optimized Cs$_2$AgIn$_{x}$Bi$_{1-x}$Cl$_6$, and with and without atomic position relaxation. The relaxation distorts the octahedra. }
    \label{Fig11}
\end{figure}
To understand the role of the lattice on the bandgap bowing, we have carried out three hypothetical experiments on Cs$_2$AgIn$_{x}$Bi$_{1-x}$Cl$_6$. (I) Across the concentration range, the lattice parameter of the doped system is taken as that of Cs$_2$AgInCl$_6$ (10.560 \AA), and band structure is calculated without relaxing the system. The resulting bandgap as a function of concentration $x$ is shown in Fig. \ref{Fig11} (a) (cyan solid line). This is a case of increasing compression with decreasing $x$. We find that, in this case, there is minimal variation in the bandgap. (II) The bandgap as a function of $x$ is now calculated by taking the lattice parameter as that of  Cs$_2$AgBiCl$_6$ (10.936 \AA). This is a case of increasing expansion with increasing $x$. The bandgap falls sharply as $x$ reaches 0.5 and then remains almost unchanged. Together these two experiments imply that (see Fig. \ref{Fig10} (a)), (i) the presence of In-$s$ orbital pushes down the Bi-$p$ states in the Bi-rich system, and (ii) in the In-rich system, the expansion significantly reduces the bandgap. (iii) In the third experiment, we carried out volume optimization and calculated the bandgap with and without structural relaxation. The relaxation primarily distorts the octahedra in this system. The results are plotted in Fig. \ref{Fig11} (b). We find the bowing is larger when the octahedra symmetry is maintained, while it reduces when the octahedra are distorted. \par
\begin{figure}[ht]
\centering
\includegraphics[angle=-0.0,scale=0.47]{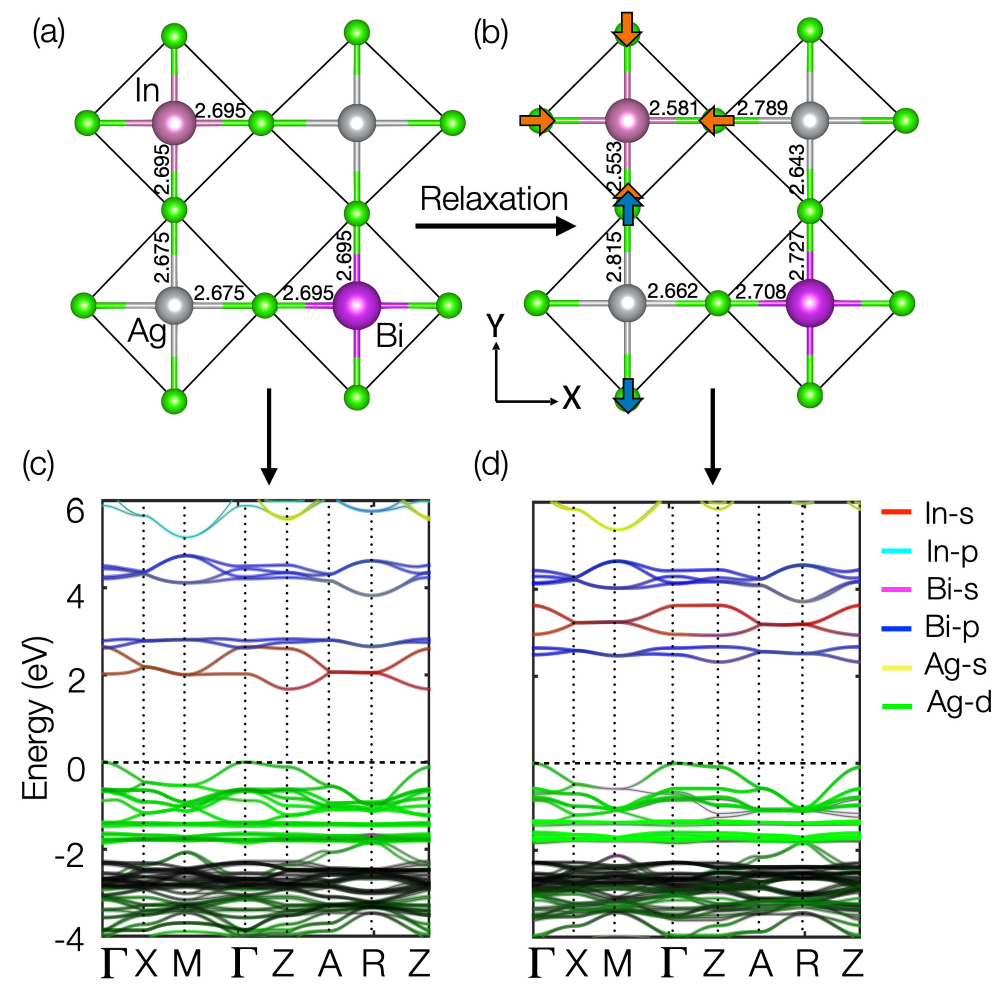}
\caption{Volume optimized Cs$_2$AgIn$_{0.5}$Bi$_{0.5}$Cl$_6$ crystal structure (a) without and (b) with  the relaxation of the atomic position. The relaxation distorts the octahedra. The orbital resolved band structures of (a) and (b) are shown in (c) and (d), respectively. }
\label{Fig12}
\end{figure}

For further analyzing the role of octahedral distortion on the bandgap, we have calculated the orbital resolved DFT band structures  of Cs$_2$AgIn$_{0.5}$Bi$_{0.5}$Cl$_6$ with and without octahedra distortion. The distortion of the octahedra includes the compression of InCl$_6$ and expansion of AgCl$_6$, as shown in Fig. \ref{Fig12}. The compression strengthens the hybridization between the In-\{$s$, $p$\} and Cl-$p$ orbitals, and therefore corresponding antibonding states ($\sigma_{s-p}^*$, $\sigma_{p-p}^*$, and $\pi_{p-p}^*$ ) go higher in energy. Since the BiCl$_6$ octahedra remain largely unaffected, the position of Bi-$p$ dominated antibonding states (blue curves) is less perturbed. This leads to a swap in the CBE character and an increase in the bandgap from 1.67 eV to 2.32 eV with distortion (as can be seen in Fig. \ref{Fig12}), and as a consequence, the bandgap bowing is weakened, which is in agreement with Fig. \ref{Fig11} (b). \par
Our overall analysis implies that the chemical effect and lattice expansion tend to increase the bandgap bowing while octahedral distortion and lattice compression reduce it. In the case of Cs$_2$Na$_{1-x}$Ag$_x$InCl$_6$ the bandgap bowing is primarily driven by the chemical effect as the lattice parameter variation is minimal. Here, the reduction of the bandgap is driven by the appearance of the Ag-$d$ states as much above the Cl-$p$ states as shown schematically in Fig. \ref{Fig10} (b).\par
\section{Conclusion and outlook}
In conclusion, we carried out a comprehensive electronic structure study by employing density functional calculations, model Hamiltonian formulation, and optoelectronic study by estimating the momentum matrix elements (MME). From our results and  analysis, we developed a theoretical workflow  to study the electronic and optoelectronic properties of halide double perovskites (pristine and doped)  for photovoltaic applications. In this work, we devise band projected molecular orbital picture (B-MOP) as an efficient tool to analyze the electronic structure of covalent hybridized systems in general and halide double perovskites (HDPs)  in particular.  Based on our understanding of the electronic structure, we could successfully categorize the HDPs into five categories which are based on the valence electron configuration of B and B$^{\prime}$, characters of the valence and conduction band edge states,  and bandgap as well as optical transition behavior. The list is summarized in Table \ref{T2}.  \par
\begin{table}[ht]
\centering
\caption {Summary of different categories (valence electron configurations) in HDPs and their electronic and optical behaviors. }
\footnotesize
\begin{tabular}{|c|c|c|c|c|c|} 
\hline
Category & B & B$^\prime$ & VB & CB &	Nature \\
\hline
$d^{10}s^0$(B)  & Group 11 & Group 15 & Ag-e$_g$ & Bi/Sb-$p$ & indirect,  \\
 - $d^{10}s^2$(B$^\prime$)& (eg. Ag) & (eg. Bi, Sb) &  & & allowed \\
\hline
$d^{10}s^0$  & Group 11 & Group 13 & Ag-e$_g$ & In/Tl-$s$ & direct,  \\
 - $d^{10}s^0$& (eg. Ag) & (eg. In, Tl) &  & & forbidden \\
\hline
$d^{10}s^2$  & Group 13 & Group 15 & In/Tl-$s$ & Bi/Sb-p & direct,  \\
 - $d^{10}s^2$& (eg. In, Tl) & (eg. Bi, Sb) &  & & allowed \\
\hline
$s^0$  & Group 1 & Group 13 & X-$p$ & In/Tl-$s$ & direct,  \\
- $d^{10}s^0$ & (eg. Na, K) & (eg. In, Tl) &  & & forbidden \\
\hline
$s^0$ - $s^2$   & Group 1 & Group 15 & Bi/Sb-$s$ & Bi/Sb-$p$ & indirect,  \\
& (eg. Na, K) & (eg. Bi, Sb) &  & & allowed \\
  &  &  &  & & (weak) \\
\hline
\end{tabular}
\label{T2}
\end{table}
The B-MOP obtained from nearest-neighbor cation-anion interactions determines the position and orbital character of the bands. The tight-binding model, which is based on second-neighbor cation-cation interactions, provides insight to determine shape and width of band dispersion.  Our study suggests that second-neighbor cation-cation interactions turn out to be deterministic factors for MME and, thereby, optoelectronic properties of HDPs. \par
Through our design principle, we show the possibilities of tuning the bandgap and optical absorption by doping. To demonstrate, we took two prototype examples of doping at the cationic site: Cs$_2$AgIn$_x$Bi$_{1-x}$Cl$_6$ and Cs$_2$Na$_{1-x}$Ag$_x$InCl$_6$ for $x$ = 0.25, 0.5, and 0.75. We obtain the maximum transition dipole moment at $x$ = 0.75 for Cs$_2$AgIn$_x$Bi$_{1-x}$Cl$_6$ and at $x$ = 0.5 for  Cs$_2$Na$_{1-x}$Ag$_x$InCl$_6$ which are found to be in good agreement with the previous experimental finding. Our analysis also provides an interesting insight into the bandgap gap bowing, which seems to be a common occurrence in HDPs. We show that the chemical effect plays an enhancement factor in bandgap bowing, while the octahedra distortion tends to minimize it.\par
The findings of the present study provide guiding principles to design efficient optoelectronic materials out of HDPs. We hope that these findings will stimulate theoretical and experimental research on alloying of HDPs to realize their applications in the area of photovoltaic and optoelectronics. The tight-binding model developed in this work (for pristine and cation intermixed) is generic and can be employed to study the electronic structure and optical behavior of 3D HDPs and their 2D and 1D counterparts \cite{ji2021}, including vacancy-induced perovskites.
\section{Acknowledgment}
This work is funded by the Department of Science and Technology (DST), India, through the Department of Science and Technology (DST) mission innovation for funding through grant number DST/TMD/IC-MAP/2K20/03 (C). We acknowledge HPCE IIT Madras for providing the computational facility.


%

\newpage

\begin{widetext}

\newcommand{\red}{\textcolor{red}}
\newcommand{\green}{\textcolor{green}}
\newcommand{\blue}{\textcolor{blue}}

\renewcommand{\thetable}{S\arabic{table}}
\renewcommand{\thefigure}{S\arabic{figure}}

\newcommand\x{\times}
\newcommand\bigzero{\makebox(0,0){\text{\huge0}}}
\newcommand*{\bord}{\multicolumn{1}{c|}{}}
\renewcommand{\figurename}{Fig.}
\renewcommand{\tablename}{TABLE}

\begin{center}
\textbf{
Supplementary Information for ``Electronic structure and optoelectronic properties of halide double perovskites:  Fundamental insights and design of a theoretical workflow"}
    
\end{center}


\section{Symmetry analysis}
As mentioned in the manuscript, all of the halide double perovskites (HDPs) adopt a cubic crystal structure featuring double space group symmetry of Fm$\bar{3}$m (no. 225). Within this symmetry group, optical transitions occur along the high-symmetry points of the Brillouin zone (BZ), namely $\Gamma$, X, L, and W (see Fig. 2(b) in the main manuscript). These high-symmetry points correspond to specific point groups such as $O_h$, $D_{3d}$, $D_{2d}$, $D_{4h}$ respectively. The character tables for these point groups are provided in the following tables (Tables \ref{tab:1} to \ref{tab:4}). Analyzing these tables will facilitate our discussion of allowed and forbidden optical transitions from a group theory perspective.
\renewcommand{\thetable}{S1}
\begin{table}[h!]
    \centering
    \caption{Character table for the double point group D$_{2d}$ at symmetry point W ((0.25, 0.5, 0.75): w.r.t. primitive vector; (0.0, 0.5, 1.0): w.r.t. conventional vectors).}
    \begin{tabular}{c|ccccc|c}
    $D_{2d}$ &E & $C_2$&$2C^\prime_2$ &$2S_4$ &$2\sigma_d$& Functions\\
    \hline
    \hline
      $W_1$& 1& 1& 1&1   &1 & $x^2 + y^2, z^2, xyz$\\
      $W_2$& 1& 1& -1&1   &-1 & $R_z, z(x^2-y^2)$\\
      $W_3$& 1& 1& 1&-1   &-1 & $x^2 - y^2$\\
      $W_4$& 1& 1& -1&-1   &1 & $z, xy, (x^2 + y^2)z, z^3$ \\
      $W_5$& 2& -2& 0&0   &0  &\begin{tabular}{@{}c@{}}$(x, y), (R_x, R_y )$\\$ (zx, yz), {x(x^2 + y^2), y(x^2 + y^2)}, (xz^2, yz^2)$,\\
      ${x(x^2 - 3y^2), y(3x^2 - y^2)}$ \end{tabular}\\
      \hline
      $W_6$& 2& 0& 0&$\sqrt{2}$   &0 & \\
      $W_7$& 2& 0& 0&-$\sqrt{2}$   &0 & \\
    \end{tabular}
    \label{tab:1}
\end{table}
\renewcommand{\thetable}{S2}
\begin{table}[h!]
    \centering
    \caption{Character table for the double point group D$_{3d}$ at symmetry point L ((0.0, 0.0, 0.0): w.r.t primitive vector; (0.5, 0.5, 0.5): w.r.t. conventional vectors.)}
    \begin{tabular}{c|cccccc|c}
         $D_{3d}$ &E & $2C_3$&$3C^\prime_2$ &$i$ &$2S_6$ &3$\sigma_d$&Functions\\
    \hline
    \hline
         $L_1^+$& 1& 1& 1&1 & 1  &1 & $x^2 + y^2, z^2$  \\ 
         $L_2^+$& 1& 1& -1&1 & 1  &-1 & $R_z$ \\
         $L_3^+$& 2& -1& 0&2 & -1  &0  & $(R_x, R_y ) (xy, x^2 - y^2), (zx, yz)$\\
         $L_1^-$& 1& 1& 1&-1 & -1  &-1 & $x(x^2 - 3y^2)$ \\
         $L_2^-$& 1& 1& -1&-1 & -1  &1 & $z$ \\
         $L_3^-$& 2& -1& 0&-2 & 1  &0 & $(x, y)$ \\
         \hline
         $L_4^+$& 2& 1& 0&2 & 1  &0  &\\
         $L_5^+$& 1& -1& $i$&1 & -1  &-$i$&  \\
         $L_6^+$& 1& -1& -$i$&1 & -1  &-$i$& \\
         $L_4^-$& 2& 1& 0&-2 & 1  &0  &\\
         $L_5^-$& 1& -1& $i$&-1 & 1  &-$i$& \\
         $L_6^-$& 1& -1& -$i$&-1 & 1  &$i$& \\
    \end{tabular}
    \label{tab:2}
\end{table}
\renewcommand{\thetable}{S3}
\begin{table}[!ht]
    \centering
    \caption{Character table for the double point Group O$_h$ at point $\Gamma$ ((0.0, 0.0, 0.0): w.r.t primitive and conventional vectors.)}
    \begin{tabular}{c|cccccccccc|c}
        $O_h$& $E$ &  $3C_2$  & $8C_3$ & $6C_4$ & $6C_2^\prime$ & $i$ & $3\sigma$&$8S_6$ & $6S_4$ & $6\sigma_d$& Functions \\
        \hline
        \hline
        $\Gamma_1^+$ & 1 & 1 & 1 & 1 & 1 & 1 & 1 & 1 & 1&1 & $x^2+y^2+z^2$\\
        $\Gamma_2^+$ & 1 & 1 & 1 & -1 & -1 & 1 & 1 & 1 & -1&-1&\\
        $\Gamma_3^+$ & 2 & 2 & -1 & 0 & 0 & 2 & 2 & -1 & 0&0& ($x^2-y^2, 2z^2-x^2-y^2$)\\
        $\Gamma_4^+$ & 3 & -1 & 0 & 1 & -1 & 3 & -1 & 0 & 1&-1& ($R_x, R_y, R_z$)\\
        $\Gamma_5^+$ & 3 & -1 & 0 & -1 & 1 & 3 & -1 & 0 & -1&1&$(xy, yz, zx)$\\
        $\Gamma_1^-$ & 1 & 1 & 1 & 1 & 1 & -1 & -1 & -1 & -1&-1& \\
        $\Gamma_2^-$ & 1 & 1 & 1 & -1 & -1 & -1 & -1 & -1 & 1&1&$xyz$\\
        $\Gamma_3^-$ & 2 & 2 & -1 & 0 & 0 & -2 & -2 & 1 &0 & 0& \\
        $\Gamma_4^-$ & 3 & -1 & 0 & 1 & -1 & -3 & 1 & 0& -1 & 1 & \begin{tabular}{@{}c@{}} $(x, y, z)$\\$(x^3, y^3, z^3)$,\\
${x(y^2 + z^2), y(z^2 + x^2), z(x^2 + y^2)}$\\
 \end{tabular}\\
        $\Gamma_5^-$ & 3 & -1 & 0 & -1 & 1 & -3 & 1 & 0& 1 & -1 &${x(z^2 - y^2), y(x^2 - z^2), z(y^2 - x^2)}$\\
        \hline
        $\Gamma_6^+$ & 2 & 0 & 1 & $\sqrt{2}$ & 0 & 2 & 0 & 1 & $\sqrt{2}$&0 &\\
        $\Gamma_7^+$ & 2 & 0 & 1 & -$\sqrt{2}$ & 0 & 2 & 0 & 1 & -$\sqrt{2}$&0 &\\
        $\Gamma_8^+$ & 4 & 0 & -1 & 0 & 0 & 
        4& 0 & -1 & 0&0 &\\
        $\Gamma_6^-$ & 2 & 0 & 1 & $\sqrt{2}$ & 0 & -2 & 0 & -1 & -$\sqrt{2}$&0 &\\
        $\Gamma_7^-$ & 2 & 0 & 1 & -$\sqrt{2}$ & 0 & -2 & 0 & -1 & $\sqrt{2}$&0 &\\
        $\Gamma_8^-$ & 4 & 0 & -1 & 0 & 0 & 
        -4& 0 & 1 & 0&0 &\\
    \end{tabular}
    \label{tab:3}
\end{table}
\renewcommand{\thetable}{S4}
\begin{table}[!ht]
    \centering
    \caption{Character Table for the double point group $D_{4h}$ at point X ((0.0, 0.5, 5.0): w.r.t primitive vector; (0.0, 0.0, 1.0): w.r.t. conventional vectors.)}
    \begin{tabular}{c|cccccccccc|c}
        $D_{4h}$& $E$ &  $2C_4$  & $C_2$ & $2C_2^\prime$ & $2C_2^{\prime\prime}$ & $i$ & $2S_4$&$ \sigma_h$ & 2$\sigma_v$ & $ 2\sigma_d$ & Functions\\
        \hline
        \hline
        $X_1^+$ & 1 & 1 & 1 & 1 & 1 & 1 & 1 & 1 & 1&1& $x^2 + y^2, z^2$\\
        $X_2^+$ & 1 & 1 & 1 & -1 & -1 & 1 & 1 & 1 & -1&-1&$R_z$\\
        $X_3^+$ & 1 & -1 & 1 & 1 & -1 & 1 & -1 & 1 & 1&-1&$x^2-y^2$\\
        $X_4^+$ & 1 & -1 & 1 & -1 & 1 & 1 & -1 & 1 & -1&1&$xy$\\
        $X_5^+$ & 2 & 0 & -2 & 0 & 0 & 2 & 0 & -2 & 0&0&$(R_x, R_y), (zx, yz)$\\
        $X_1^-$ & 1 & 1 & 1 & 1 & 1 & -1 & -1 & -1 & -1&-1&\\
        $X_2^-$ & 1 & 1 & 1 & -1 & -1 & -1 & -1 & -1 & 1&1& $z, (x^2+y^2)z, z^3$\\
        $X_3^-$ & 1 & -1 & 1 & 1 & -1 & -1 & 1 & -1 & -1&1&$xyz$\\
        $X_4^-$ & 3 & -1 & 0 & 1 & -1 & -3 & 1 & 0& -1 & 1&$z(x^2-y^2)$\\
        $X_5^-$ & 2 & 0 & -2 & 0 & 0 & -2 & 0 & 2 & 0&0&\begin{tabular}{@{}c@{}}$(x, y)$\\
        ${x(x^2+y^2 ), y(x^2+y^2)}, (xz^2,yz^2 ),$\\ ${x(x^2-3y^2), y(3x^2-y^2)}$ \end{tabular}\\
        \hline
        $X_6^+$ & 2 & $\sqrt{2}$ & 0 & 0 & 0 & 2 & $\sqrt{2}$ & 0 & 0 &0\\
        $X_7^+$ & 2 & -$\sqrt{2}$ & 0 & 0 & 0 & 2 & $\sqrt{2}$ & 0 & 0 &0\\
        $X_6^-$ & 2 & $\sqrt{2}$ & 0 & 0 & 0 & -2 & -$\sqrt{2}$ & 0 & 0 &0\\
        $X_7^-$ & 2 & -$\sqrt{2}$ & 0 & 0 & 0 & -2 & $\sqrt{2}$ & 0 & 0 &0\\
    \end{tabular}
    \label{tab:4}
\end{table}

\subsection{Band-diagram analysis}
The irreducible representations (IR) at the band edges can be explained via group theoretical analysis. As an example, for Cs$_2$AgBiCl$_6$ the conduction band and the valence band primarily arise from the contributions from  Ag$-e_g$, Bi$-p, s$ and Cl$-p$ states. Consequently, these bands can be expressed as a linear combination of $|j,m\rangle$ states like - $|\frac{1}{2}, \frac{1}{2}\rangle$, $|\frac{3}{2}, \frac{1}{2}\rangle$, $|\frac{5}{2}, \frac{3}{2}\rangle$. By closely analyzing the band structure, for the symmetry point W and the corresponding D$_{2d}$ point group, we can deduce that only $W_6$ and $W_7$ represent these kind of orbital combinations such as - $\phi (|\frac{1}{2}, \frac{1}{2}\rangle, |\frac{1}{2}, \bar{\frac{1}{2}}\rangle)$; $\phi(|\frac{3}{2},\frac{1}{2}\rangle, -|\frac{3}{2}, \bar{\frac{3}{2}}\rangle)$; $\phi(|\frac{3}{2}, \bar{\frac{3}{2}}\rangle, -|\frac{3}{2}, \frac{3}{2}\rangle)$; $\phi(|\frac{5}{2}, \bar{\frac{3}{2}}\rangle$, $|\frac{5}{2}, \frac{3}{2}\rangle)$ and many more. Likewise, the IRs at other symmetry points L ($L_5^-$, $L_4^-$), $\Gamma$ ($\Gamma_8^+$, $\Gamma_6^-$), X ($X_6^+$, $X_6^-$) also represent similar orbital combinations which are expected from our calculations. Therefore, the IRs of the band diagrams obtained from DFT agree well with our group theoretical analysis.\\

\section{Explanation of Optical Transitions from Group Theory}
Here, we are going to discuss the optical transitions primarily driven by the direct products of IRs as well as the parity of these representations. For a comprehensive discussion of these transitions, we have written the direct product tables (Table \ref{tab:5} - \ref{tab:8}) for the aforementioned point groups at their corresponding high symmetry points.\\
\renewcommand{\thetable}{S5}
\begin{table}[h!]
    \centering
    \caption{Direct product table for group $D_{2d}$.}
    \begin{tabular}{c|ccccccc}
         $D_{2d}$ &$W_1$ & $W_2$ & $W_3$ &$W_4$ &$W_5$ & $W_6$ & $W_7$\\
    \hline
    \hline
      $W_1$ & $W_1$ & $W_2$& $W_3$& $W_4$  & $W_5$ & $W_6$ & $W_7$\\
      $W_2$& & $W_2$& $W_3$  & $W_4$ & $W_5$ & $W_6$ & $W_7$  \\
      $W_3$& & &$W_1$  &$W_2$ & $W_5$ & $W_7$ & $W_6$  \\
      $W_4$& & &  &$W_1$ & $W_5$ & $W_7$ & $W_6$  \\
      $W_5$& & & & & $W_1$+$W_2$+$W_3$+$W_4$ & $W_6$+$W_7$ & $W_6$+$W_7$ \\
      $W_6$& & & & & & $W_1$+$W_2$+$W_5$ & $W_3$+$W_4$+$W_5$  \\
      $W_7$& & & & & & & $W_1$+$W_2$+$W_5$  \\     
    \end{tabular}
    \label{tab:5}
\end{table}
\renewcommand{\thetable}{S6}
\begin{table}[h!]
    \centering
    \caption{Direct product table for group $D_{3d}$.}
    \begin{tabular}{c|cccccc}
         $D_{3d}$ &$L_4^+$ & $L_5^+$ & $L_6^+$ &$L_4^-$ &$L_5^-$ & $L_6^-$ \\
         \hline
         \hline
         $L_1^+$ & $L_4^+$ & $L_5^+$ & $L_6^+$ &$L_4^-$ &$L_5^-$ & $L_6^-$\\
         $L_2^+$ & $L_4^+$ & $L_6^+$ & $L_5^+$ &$L_4^-$ &$L_6^-$ & $L_5^-$\\
         $L_3^+$ & $L_4^+$+$L_5^+$+$L_6^+$ & $L_4^+$ & $L_4^+$ &$L_4^-$+$L_5^-$+$L_6^-$&$L_4^-$ & $L_4^-$\\
         $L_1^-$ & $L_4^-$ & $L_5^-$ & $L_6^-$ &$L_4^+$ &$L_5^+$ & $L_6^+$\\
         $L_2^-$ & $L_4^-$ & $L_6^-$ & $L_5^-$ &$L_4^+$ &$L_6^+$ & $L_5^+$\\
         $L_3^-$ & $L_4^-$+$L_5^-$+$L_6^-$ & $L_4^-$ & $L_4^-$ &$L_4^+$+$L_5^+$+$L_6^+$ &$L_4^+$ & $L_4^+$\\
         $L_5^+$ & $L_1^+$+$L_2^+$+$L_3^+$ & $L_3^+$ & $L_3^+$ & $L_1^-$+$L_2^-$+$L_3^-$ &$L_3^-$ & $L_3^-$\\
         $L_6^+$ &  & $L_2^+$ & $L_1^+$ &$L_3^-$ &$L_2^-$ & $L_1^-$\\
         $L_7^+$ &  &  & $L_2^+$ &$L_3^-$ &$L_1^-$ & $L_2^-$\\
         $L_5^-$ &  &  &  & $L_1^+$+$L_2^+$+$L_3^+$ &$L_3^+$ & $L_3^+$\\
         $L_6^-$ & & & & &$L_2^+$ & $L_1^+$\\
         $L_7^-$ & & & & & & $L_2^+$\\  
    \end{tabular}
    \label{tab:6}
\end{table}
\renewcommand{\thetable}{S7}
\begin{table}[h!]
    \centering
    \caption{Direct product for group $D_{4h}$}
    \begin{tabular}{c|cccc}
        $D_{4h}$ &$X_6^+$ & $X_7^+$ & $X_6^-$ &$X_7^-$ \\
        \hline
        \hline
        $X_1^+$ & $X_6^+$ & $X_7^+$ & $X_6^-$ &$X_7^-$ \\
        $X_2^+$ & $X_6^+$ & $X_7^+$ & $X_6^-$ &$X_7^-$ \\
        $X_3^+$ & $X_7^+$ & $X_6^+$ & $X_7^-$ &$X_6^-$ \\
        $X_4^+$ & $X_7^+$ & $X_6^+$ & $X_7^-$ &$X_6^-$ \\
        $X_5^+$ & $X_6^+$+$X_7^+$ & $X_6^+$+$X_7^+$ & $X_6^-$+$X_7^-$ &$X_6^-$+$X_7^-$ \\
        $X_1^-$ & $X_6^-$ & $X_7^-$ & $X_6^+$ &$X_7^+$ \\
        $X_2^-$ & $X_6^-$ & $X_7^-$ & $X_6^+$ &$X_7^+$ \\
        $X_3^-$ & $X_7^-$ & $X_6^-$ & $X_7^+$ &$X_6^+$ \\
        $X_4^-$ & $X_7^-$ & $X_6^-$ & $X_7^+$ &$X_6^+$ \\
        $X_5^-$ & $X_6^-$+$X_7^-$ & $X_6^-$+$X_7^-$ & $X_6^+$+$X_7^+$ &$X_6^+$+$X_7^+$ \\
        $X_6^+$ & $X_1^+$+$X_2^+$+$X_5^+$ & $X_3^+$+$X_4^+$+$X_5^+$ & $X_1^-$+$X_2^-$+$X_5^-$ &$X_3^-+$+$X_4^-$+$X_5^-$ \\
        $X_7^+$ &  &$X_1^+$+$X_2^+$+$X_5^+$ & $X_1^-$+$X_2^-$+$X_5^-$ &$X_3^-+$+$X_4^-$+$X_5^-$ \\
        $X_6^-$ &  &  &  $X_1^+$+$X_2^+$+$X_5^+$ & $X_3^+$+$X_4^+$+$X_5^+$ \\
        $X_7^-$ &  &  &  & $X_1^+$+$X_2^+$+$X_5^+$ \\
    \end{tabular}
    \label{tab:7}
\end{table} 
\renewcommand{\thetable}{S8}
\begin{table}[h!]
\hspace{-2cm}
\caption{Direct product for group $O_{h}$}
    \begin{tabular}{c|cccccc}
    $O_h$ & $\Gamma_6^+$ & $\Gamma_7^+$ & $\Gamma_8^+$ & $\Gamma_6^-$ & $\Gamma_7^-$ & $\Gamma_8^-$\\
     \hline
     \hline
     $\Gamma_6^+$ & $\Gamma_1^+$+ $\Gamma_4^+$ & $\Gamma_2^+$+$\Gamma_5^+$ & $\Gamma_3^+$+$\Gamma_4^+$+$\Gamma_5^+$ & $\Gamma_1^-$+ $\Gamma_4^-$ & $\Gamma_2^-$+$\Gamma_5^-$ & $\Gamma_3^-$+$\Gamma_4^-$+$\Gamma_5^-$\\
     $\Gamma_7^+$ &  & $\Gamma_1^+$+ $\Gamma_4^+$  & $\Gamma_3^+$+$\Gamma_4^+$+$\Gamma_5^+$ & $\Gamma_2^-$+$\Gamma_5^-$ &$\Gamma_1^-$+ $\Gamma_4^-$ & $\Gamma_3^-$+$\Gamma_4^-$+$\Gamma_5^-$\\
     $\Gamma_8^+$ &  &   & $\Gamma_1^+$+ $\Gamma_2^+$+$\Gamma_3^+$+2$\Gamma_4^+$+2$\Gamma_5^+$ & $\Gamma_3^-$+$\Gamma_4^-$+$\Gamma_5^-$ &$\Gamma_3^-$+$\Gamma_4^-$+$\Gamma_5^-$ & $\Gamma_1^-$+ $\Gamma_2^-$+$\Gamma_3^-$+2$\Gamma_4^-$+2$\Gamma_5^-$\\
     $\Gamma_6^-$ &  &   & & $\Gamma_1^+$+ $\Gamma_4^+$ & $\Gamma_2^+$+ $\Gamma_5^+$&$\Gamma_3^-$+$\Gamma_4^-$+$\Gamma_5^+$\\
     $\Gamma_7^-$ &  &   & &  & $\Gamma_1^+$+ $\Gamma_4^+$ &$\Gamma_3^-$+$\Gamma_4^-$+$\Gamma_5^+$\\
     $\Gamma_7^-$ &  &   & &  &  & $\Gamma_1^+$+ $\Gamma_2^+$+$\Gamma_3^+$+$2\Gamma_4^+$+$2\Gamma_5^+$\\
    \end{tabular}
    \label{tab:8}
\end{table}

\subsection{Direct transitions} The direct optical transitions between VBE and CBE states  are  guided by two key rules:\\
(i) For a non-trivial transition, the direct product of the involved states must exhibit electric dipole characteristics consistent with the symmetries of their respective point groups.\\
(ii) Transitions between the two states are permissible only if they possess opposite parities, in accordance with Laporte's rule \cite{Laporte:25}.\\
Table \ref{tab:9} lists the distinct types of direct transitions that are pertinent to various HDP structures along the high-symmetry points. As an example, we can consider the direct transitions within the Cs$_2$AgBiCl$_6$ structure. At point W, there is only one possible transition from $W_6$ to $W_7$ between VBE and CBE. 
From the Table \ref{tab:9} we note that the direct product between $W_6$ and $W_7$ is expressed as the summation of three representations $W_3$, $W_4$ and $W_5$. Among these three representations, $W_4$ has the character of electric dipole operator (A$_2$) for D$_{2d}$ point group. Hence, this transition is allowed. 
At L, the direct product between $L_5^-+L_6^-$ and $L_4^-$ can be expressed as  $L_1^+$+$L_2^+$+$L_3^+$, which does not include the electric dipole operator $L_2^-$ (A$_{1u}$)+$L_3^-$(E$_u$) for the point group D$_{3d}$. Hence, This transition is forbidden which is also supported by the Laporte rule as well as by the P$^2$ (see Fig. 6 of the main manuscript). Again, the transitions between VBE and CBE along $\Gamma$ and X point are also allowed ((Table \ref{tab:9})) following the direct product rule as demonstrated in Table \ref{tab:8} and \ref{tab:7} respectively.
\renewcommand{\thetable}{S9}
\begin{table}[h!]
    \centering
    \caption{Direct products between two IRs for all types of allowed transitions at high symmetry points in the HDP systems. }
    \begin{tabular}{cc}
    \hline
    \hline
       \textbf{ W:} & \begin{tabular}{@{}c@{}}$W_6 \otimes W_7$ = $W_2 \oplus W_3 \oplus W_5$ \\
$W_6 \otimes W_6$ = $W_1 \oplus W_4 \oplus W_5$ \\
$W_7 \otimes W_7$ = $W_1 \oplus W_4 \oplus W_5$ \end{tabular} \\
\hline
       $\bf{\Gamma:}$  & \begin{tabular}{@{}c@{}} $\Gamma_8^+ \otimes \Gamma_6^-$ = $\Gamma_3^- \oplus \Gamma_4^- \oplus \Gamma_5^-$ \\
$\Gamma_6^+ \otimes \Gamma_6^-$ = $\Gamma_1^- \oplus \Gamma_4^-$ \end{tabular} \\
\hline
\textbf{L:} & $L_4^+ \otimes L_4^-$ = $L_1^- \oplus L_2^- \oplus L_3^-$ \\
\hline
\textbf{X:} & $X_6^+ \otimes X_6^-$ = $X_1^- \oplus X_2^- \oplus X_5^-$\\
\hline
\hline
    \end{tabular}
    \label{tab:9}
\end{table}

 \subsection{Indirect Transitions} 
 As we can observe from Fig. 6 of the main manuscript there are only two indirect bandgap semiconductors 
 Cs$_2$AgBiCl$_6$ and Cs$_2$AgSbCl$_6$ which represent finite optical transition along their band spectra. Given that indirect semiconductors primarily involve transitions between VBM and CBM via indirect processes, our discussion on selection rules will specifically pertain to these two cases only.
 As evident from the dispersion spectrum (Fig. 6 in the main manuscript), there are only two specific types of indirect transitions, namely $W_6 \rightarrow L_4^-$ and $X_6^+ \rightarrow L_4^-$. 
 Since indirect transitions are mainly phonon-mediated interactions, we need to analyze the active phonon modes for their respective point groups. As determined from the character table of D$_{2d}$ and D$_{3d}$ for W and X point respectively, it can be written that $W_6 \otimes L_4^-$ = $W_1 \oplus W_2 \oplus W_3 \oplus W_4$. Now, $W_1$ and $W_3$ have the optical phonon, and $W_4$ has the acoustic phonon characteristics. Therefore, the transition $W_6 \rightarrow L_4^-$ is a phonon-mediated allowed transition.
Similarly, $X_6^+ \otimes L_4^-$ = $X_1^- \oplus X_2^- \oplus X_5^-$. This transition is also allowed only through the acoustic phonon modes $X_2^-$ and $X_5^-$.
Likewise, indirect transitions in other HDPs can be successfully demonstrated through an elementary group theoretical approach.

\section{Tight-Binding model Hamiltonian}
\renewcommand{\thefigure}{S1}
\begin{figure}[h]
    \centering
    \includegraphics[scale=0.4]{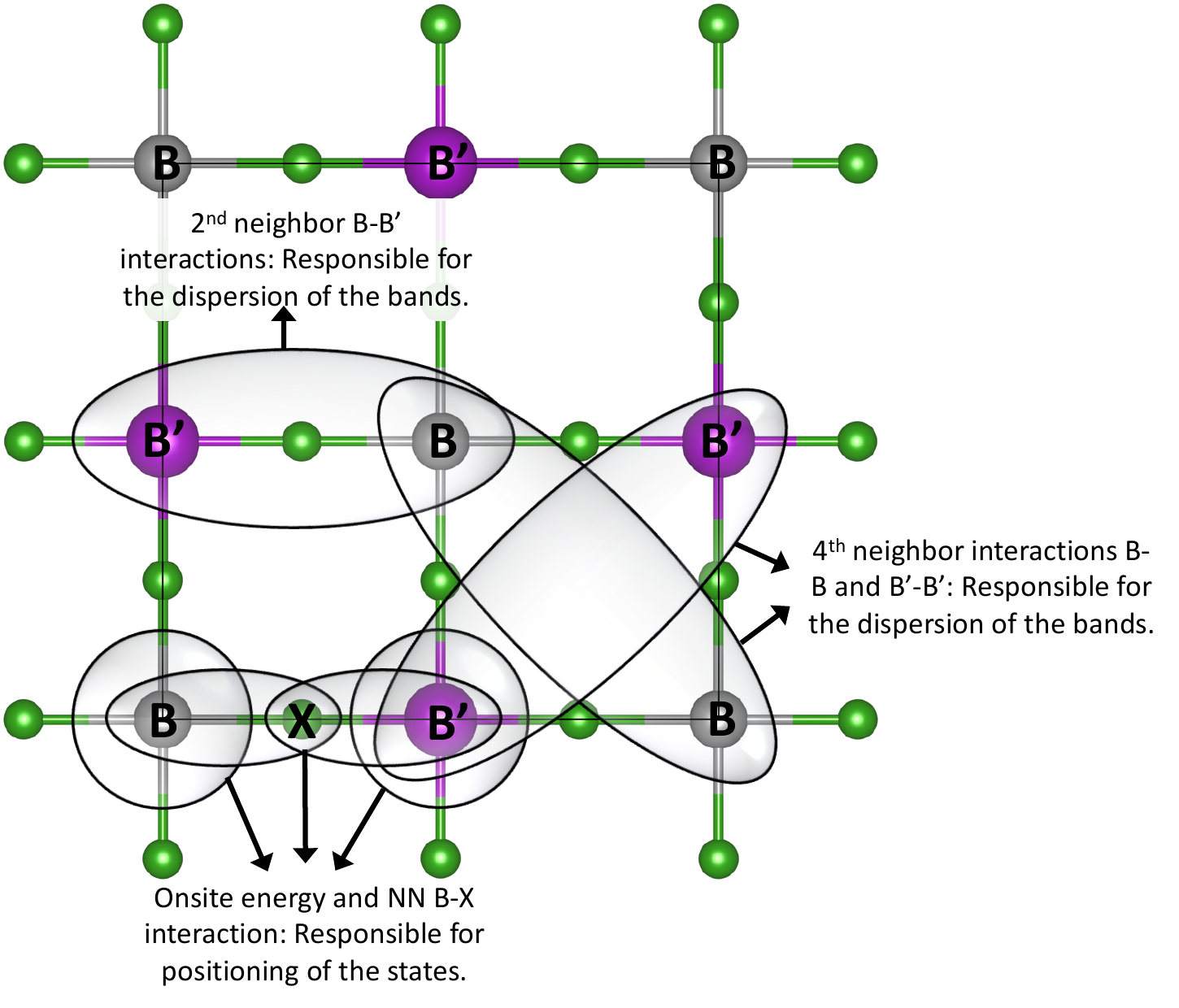}
    \caption{Schematic illustration of various hopping interactions up to the fourth nearest neighbor (NN) in halide double perovskites. The Hamiltonian presented below involves effective interactions among the B and B$^\prime$ orbitals to describe the band structure in the vicinity of Fermi level. The contribution of NN B-X and B$^\prime$-X interactions is incorporated in these effective interactions. It can be understood by examining the downfold technique such as Lo\"wdin downfold \cite{Lowdin, mayank-2022}. }
    \label{TB-interactions}
\end{figure}
The generic tight-binding (TB) Hamiltonian incorporating covalent interactions, as illustrated in Fig. \ref{TB-interactions}, and spin-orbit coupling (SOC) effect for HDP  can be written as
\begin{equation}
H = \sum_{i, \alpha} \epsilon_{i\alpha} c^\dagger_{i\alpha}c_{i\alpha} + \sum_{{\braket{ij}};\alpha,\beta} t_{i\alpha j\beta}(c^\dagger_{i\alpha}c_{j\beta}+h.c.)+\lambda \vec{L}\cdot\vec{S},
\end{equation}
The details of the Hamiltonian are provided in Section V of the main text. Here, we have explained the components of the Hamiltonian matrix. The complete Hamiltonian matrix containing interactions between B and B$^\prime$ is expressed as
\begin{equation}
    H = \left[\begin{array}{cc}
    H_{BB} & H_{BB'}\\
    H_{BB'}^\dagger&H_{B'B'}\\
    \end{array}\right]
\end{equation}
Where $H_{BB/B^\prime B^\prime}$ are the Hamiltonian sub-matrices containing interactions among the same cations (\textit{i.e.}, the onsite energy and fourth-neighbor $BB$ and $B^\prime B^\prime$ interactions) and the Hamiltonian sub-matrices $H_{BB^\prime}$ and $H_{BB^\prime}^\dagger$ contain the second neighbor $BB^\prime$ interactions between two different cations (see Fig. \ref{TB-interactions}).
\begin{equation}
    H_{BB/B'B'} = \left[\begin{array}{cc}
    H^{\uparrow\uparrow} & 0\\
    0 &H^{\downarrow\downarrow}\\
    \end{array}\right], 
        H_{BB^\prime/B^\prime B} = \left[\begin{array}{cc}
    H^{\uparrow\uparrow} & 0\\
    0 &H^{\downarrow\downarrow}\\
    \end{array}\right] 
\end{equation}
The Hamiltonian H$_{BB/B^\prime B^\prime}$ for 4$^{th}$ NN interactions, when $s$ and $p$ orbitals of B and B$^\prime$ are contributing to bands around the Fermi are in basis set order $\ket{s^{B/B^\prime,\uparrow/\downarrow}}$, $\ket{p_x^{B/B^\prime,\uparrow/\downarrow}}$, $\ket{p_y^{B/B^\prime,\uparrow/\downarrow}}$, and $\ket{p_z^{B/B^\prime,\uparrow/\downarrow}}$ is given as
\begin{equation}
    H^{\uparrow\uparrow/\downarrow\downarrow}_{BB^\prime/B^\prime B/BB/B^\prime B^\prime} =  \left[\begin{array}{ccccc}
    \epsilon_s + h_{ss} & h_{sp_x}& h_{sp_y}& h_{sp_z}\\
    h_{sp_x}^* & \epsilon_p +  h_{p_xp_x} & h_{p_xp_y} & h_{p_xp_z}\\
    h_{sp_y}^* & h_{p_xp_y}^* & \epsilon_p + h_{p_yp_y} & h_{p_yp_z}\\
    h_{sp_z}^* & h_{p_xp_z}^* & h_{p_yp_z}^* & \epsilon_p + h_{p_zp_z}\\
    \end{array}\right]
\end{equation}
This is also the Hamiltonian matrix for 2$^{nd}$ neighbor B-B$^\prime$ interactions when $s$ and $p$ orbitals of B and B$^\prime$ are contributing, however, in that case, hopping parameters will be different. \\

Furthermore, the Hamiltonian matrix for 4$^{th}$ neighbor B-B interactions, when an Ag atom is sitting at B-site, in basis set order  $\ket{s^{B,\uparrow/\downarrow}}$, $\ket{d_{z^2}^{B,\uparrow/\downarrow}}$, $\ket{d_{x^2-y^2}^{B,\uparrow/\downarrow}}$, $\ket{d_{xy}^{B,\uparrow/\downarrow}}$, $\ket{d_{xz}^{B,\uparrow/\downarrow}}$, and $\ket{d_{yz}^{B,\uparrow/\downarrow}}$, can be expressed as 
    \begin{equation}
    H^{\uparrow\uparrow/\downarrow\downarrow}_{BB} = 
    \left[\begin{array}{ccccccc}
    \epsilon_s + h_{ss} & h_{sd_{z^2}}& h_{sd_{x^2-y^2}}& h_{sd_{xy}} &  h_{sd_{xz}} &  h_{sd_{yz}}\\
    h_{sd_{z^2}}^* & \epsilon_{de_g} + h_{d_{z^2}d_{z^2}}& h_{d_{z^2}d_{x^2-y^2}}& h_{d_{z^2}d_{xy}} &  h_{d_{z^2}d_{xz}} &  h_{d_{z^2}d_{yz}}\\
    h_{sd_{x^2-y^2}}^* & h_{d_{x^2-y^2}d_{z^2}} & \epsilon_{d_{x^2-y^2}} + h_{d_{x^2-y^2}d_{x^2-y^2}}& h_{d_{x^2-y^2}d_{xy}} &  h_{d_{x^2-y^2}d_{xz}} &  h_{d_{x^2-y^2}d_{yz}}\\
    h_{sd_{xy}}^* & h_{d_{xy}d_{z^2}} &  h_{d_{x^2-y^2}d_{x^2-y^2}} & \epsilon_{dxy} +h_{d_{xy}d_{xy}} &  h_{d_{xy}d_{xz}} &  h_{d_{xy}d_{yz}}\\
    h_{sd_{xz}}^* & h_{d_{xz}d_{z^2}} &  h_{d_{xz}d_{x^2-y^2}} & h_{d_{xz}d_{xy}} & \epsilon_{dxz} +h_{d_{xz}d_{xz}} &  h_{d_{xz}d_{yz}}\\
    h_{sd_{yz}}^* & h_{d_{yz}d_{z^2}} &  h_{d_{yz}d_{x^2-y^2}} & h_{d_{yz}d_{xy}} &  h_{d_{yz}d_{xz}} & \epsilon_{dyz} + h_{d_{yz}d_{yz}}\\
    \end{array}\right]
    \end{equation}
and the Hamiltonian sub-matrix $H_{BB^\prime/B^\prime B}$, containing interactions between $d$ and $p$ orbitals from two different NN cations can be expressed as
\begin{equation}
    H^{\uparrow\uparrow/\downarrow\downarrow}_{BB^\prime} = \left[\begin{array}{cccc}
         h_{ss}& h_{sp_x} & h_{sp_y} & h_{sp_z}   \\
         h_{d_{z^2}s} & h_{d_{z^2}p_x} & h_{d_{z^2}p_y} & h_{d_{z^2}p_z} \\
         h_{d_{x^2-y^2}s} & h_{d_{x^2-y^2}p_x} & h_{d_{x^2-y^2}p_y} & h_{d_{x^2-y^2}p_z} \\
         h_{d_{xy}s} & h_{d_{xy}p_x} & h_{d_{xy}p_y} & h_{d_{xy}p_z} \\
         h_{d_{xz}s} & h_{d_{xz}p_x} & h_{d_{xz}p_y} & h_{d_{xz}p_z} \\
         h_{d_{yz}s} & h_{d_{yz}p_x} & h_{d_{yz}p_y} & h_{d_{yz}p_z} \\
    \end{array}\right].
\end{equation}
Each of the above elements of the Hamiltonian matrix $h_{ij}$ are expressed in terms of Slater-Koster \cite{Slater} form.
The SOC matrix for $d$-orbital basis set in basis order  $\ket{s^{B,\uparrow}}$, $\ket{d_{z^2}^{B,\uparrow}}$, $\ket{d_{x^2-y^2}^{B,\uparrow}}$, $\ket{d_{xy}^{B,\uparrow}}$, $\ket{d_{xz}^{B,\uparrow}}$, $\ket{d_{yz}^{B,\uparrow}}$, $\ket{s^{B,\downarrow}}$, $\ket{d_{z^2}^{B,\downarrow}}$, $\ket{d_{x^2-y^2}^{B,\downarrow}}$, $\ket{d_{xy}^{B,\downarrow}}$, $\ket{d_{xz}^{B,\downarrow}}$, $\ket{d_{yz}^{B,\downarrow}}$ and $p$-orbital basis set in the basis order $\ket{s^{B/B^\prime,\uparrow}}$, $\ket{p_x^{B/B^\prime,\uparrow}}$, $\ket{p_y^{B/B^\prime,\uparrow}}$, $\ket{p_z^{B/B^\prime,\uparrow}}$, $\ket{s^{B/B^\prime,\downarrow}}$, $\ket{p_x^{B/B^\prime,\downarrow}}$, $\ket{p_y^{B/B^\prime,\downarrow}}$, $\ket{p_z^{B/B^\prime,\downarrow}}$ are 
\begin{equation}
    H_{d}^{SOC} = \lambda \left[\begin{array}{cccccccccccc}
         0 & 0 & 0 & 0 & 0 & 0 & 0 & 0 & 0 & 0 & 0 & 0 \\
         0 & 0 & 0 & 0 & 0 & 0 & 0 & 0 & 0 & 0 & -\sqrt{3} & i\sqrt{3} \\
         0 & 0 & 0 & -2i& 0 & 0 & 0 & 0 & 0 & 0 & 1 & i\\
         0 & 0 & 2i& 0 & 0 & 0 & 0 & 0 & 0 & 0 & -i& 1 \\
         0 & 0 & 0 & 0 & 0 & -i& 0 & \sqrt{3} & -1 & i& 0 & 0 \\
         0 & 0 & 0 & 0 & i& 0 & 0 & -i\sqrt{3} & -i& -1 & 0 & 0 \\
         0 & 0 & 0 & 0 & 0 & 0 & 0 & 0 & 0 & 0 & 0 & 0 \\
         0 & 0 & 0 & 0 & \sqrt{3} & i\sqrt{3} & 0 & 0 & 0 & 0 & 0 & 0 \\
         0 & 0 & 0 & 0 & -1 & i& 0 & 0 & 0 & 2i& 0 & 0 \\
         0 & 0 & 0 & 0 & -i& -1 & 0 & 0 & -2i& 0 & 0 & 0 \\
         0 & -\sqrt{3} & 1 & i& 0 & 0 & 0 & 0 & 0 & 0 & 0 & i\\
         0 & -i\sqrt{3} & -i& 1 & 0 & 0 & 0 & 0 & 0 & 0 & -i& 0 \\
    \end{array}\right], 
    \end{equation}
    \begin{equation}
    H_{p}^{SOC} = \left[\begin{array}{cccccccc}
         0 & 0 & 0 & 0 & 0 & 0 & 0 & 0\\
         0 & 0 & -i\lambda & 0 & 0 & 0  & 0 & \lambda \\
         0 & i\lambda & 0 & 0 & 0 & 0  & 0 & -i\lambda \\
         0 & 0 & 0 & 0 & 0 & -\lambda  & i\lambda & 0 \\
         0 & 0 & 0 & 0 & 0 & 0 & 0 & 0\\
         0 & 0 & 0 & -\lambda & 0 & 0 & i\lambda & 0\\
         0 & 0 & 0 & -i\lambda & 0 & -i\lambda & 0 & 0\\
         0 & \lambda & i\lambda & 0 & 0 & 0 & 0 & 0\\
    \end{array}\right]
\end{equation}
\section{Estimation of TB hopping parameters}
TB hoppings for various HDPs obtained from fitting with DFT band structures are provided in Tables \ref{T2}, \ref{T3}, and \ref{T4}.
\renewcommand{\thetable}{S10}
\begin{small}
\begin{sidewaystable}
\caption {TB hopping parameters (in eV) for halide perovskites family.}
\begin{tabular}{ |c|c|c|c|c|c|c|c|c|c|} 
\hline
 &	hopping  &	$Cs_2AgBiCl_6$ & $Cs_2AgSbCl_6$ & $Cs_2AgAsCl_6$  & $Cs_2AgInCl_6$  & $Cs_2AgTlCl_6$ & $Cs_2AgGaCl_6$ & $Cs_2CuBiCl_6$  & $Cs_2CuInCl_6$\\
 &parameters	 &	B = Ag,B$^\prime$ = Bi& B = Ag,B$^\prime$ = Sb & B = Ag,B$^\prime$ = As  & B = Ag,B$^\prime$ = In  & B = Ag,B$^\prime$ = Tl & B = Ag,B$^\prime$ = Ga & B = Cu,B$^\prime$ = Bi& B = Cu,B$^\prime$ = In\\
\hline
&$\epsilon_s^{BB}$&   7.76 &	5.86 & 6.16 & 7.15 &		7.76  & 7.15 & 5.4 & 5.2\\
&$\epsilon_{e_g}^{BB}$& -0.40 &	-1.1 & -0.92   & -0.08 &		0.04 & -0.05 & -0.17 &-0.08\\
&$\epsilon_{t_{2g}}^{BB}$&	-0.91 &	-1.81 &  -1.61 &	-0.82 &	-1.16 & -0.76 & -0.70 & -0.55\\
B-B&$t_{dd\sigma}^{BB}$& -0.002 &	0.0& 0.0&	0.01 &	0.016 & 0.0& 0.0& 0.0\\
&$t_{dd\pi}^{BB}$&	    -0.008 &	0.0 &0.0&	0.007 &	-0.017& 0.0& 0.0& 0.0\\
&$t_{dd\delta}^{BB}$&  0.0 & 0.0 &0.0&	0.0 & 0.0& 0.0& 0.0& 0.0\\
&$t_{ss\sigma}^{BB}$& 	0.0 &	0.03 &0.0&	0.0 &	0.0& 0.0& 0.0& 0.0\\
&$t_{sd\sigma}^{BB}$&	0.0 &	0.0 &0.0&	0.0 &	0.0& 0.0& 0.0& 0.0\\
\hline
&$\epsilon_{s}^{B^\prime B^\prime}$&	-1.85 &	-1.16 & -1.75 &	4.6 &	1.46& 4.15 & -3.2 & 1.44\\
&$\epsilon_{p}^{B^\prime B^\prime}$&	4.18 &	3.41 & 3.30 &	7.98 &	6.81 & 7.40 & 2.3& 5.8\\
&$t_{ss}^{B^\prime B^\prime}$&	-0.03 &	-0.065 & -0.01& 0.0 & 0.048& 0.0 & -0.03&-0.06\\
B$^\prime$-B$^\prime$ &$t_{sp}^{B^\prime B^\prime}$&	0.03 &	0.02 &	0.03& 0.018 &	0.126 & 0.018 & 0.025&0.0\\
&$t_{pp\sigma}^{B^\prime B^\prime}$&	-0.028 &	-0.076 &-0.06&	-0.092 &	-0.078& -0.092 &-0.055& -0.25\\
&$t_{pp\pi}^{B^\prime B^\prime}$&	0.0 &	0.0 & 0.0&	0.0 &	0.0 &0.0 & -0.014 & 0.0\\
\hline
&$t_{ss}^{BB^\prime}$&	0.0 &	0.0 & 0.24&	0.375 &	0.532 &0.384 & 0.22 & 0.25\\
&$t_{sp}^{BB^\prime}$&	0.475 &	0.57 & 0.60&	0.48 &	0.54 & 0.48 & 0.57 & 0.54\\
B-B$^\prime$&$t_{sd\sigma}^{BB^\prime}$&	0.193 & 	0.25 & 0.32& 0.42 &	0.39 &0.44 & 0.12 & 0.27\\
&$t_{pd\sigma}^{BB^\prime}$&	0.54 &	0.63 & 0.67&	0.63 &	0.84 & 0.64& 0.38 & 0.22\\
&$t_{pd\pi}^{BB^\prime}$&	0.0 &	0.14 & 0.0&	0.24 &	0.2 &0.24& 0.02 & 0.12\\
\hline
SOC&$\lambda^B$&	0.09&	0.09& 0.09&	0.09&	0.09 & 0.09 & 0.0 & 0.06\\
&$\lambda^{B^\prime}$&	0.54&	0.17& 0.1&	0.16&	0.24 & 0.05 & 0.16& 0.16\\
\hline
\end{tabular}
\label{T2}
\end{sidewaystable}
\end{small}
\renewcommand{\thetable}{S11}
\begin{table*}[h]
\centering
\caption {TB hopping parameters (in eV) for halide perovskites family.}
\begin{tabular}{ |c|c|c|c|c|c|c|} 
\hline
Cation- &	TB parameters & $CsPbBr_3$ &	$Cs_2InBiCl_6$  & $Cs_2InSbCl_6$  & $Cs_2TlBiCl_6$  & $Cs_2TlSbCl_6$  \\
cation & &&	B = In, B$^\prime$ = Bi & B = In, B$^\prime$ = Sb & B = Tl, B$^\prime$ = Bi & B = Tl, B$^\prime$ = Sb \\
\hline
&$\epsilon_s^{BB}$& -1.36 &  -0.98 &	-1.16 & -1.0 & -1.08  \\
&$\epsilon_{p}^{BB}$& 4.15 & 4.5 &	4.07 &	5.24 &	4.6 \\
B-B&$t_{ss\sigma}^{BB}$& 0.0 &	-0.053 &	-0.008 &	-0.026 &	0.0 \\
&$t_{sp\sigma}^{BB}$& 0.0 &	-0.016 &	0.036 &	0.044 &	0.028 \\
&$t_{pp\sigma}^{BB}$& 0.0  &	0.0 &	0.0 &	0.04 &	0.0 \\
&$t_{pp\pi}^{BB}$& 0.0 &	0.0 &	0.0 &	0.0 &	0.0 \\
\hline
&$\epsilon_{s}^{B^\prime B^\prime}$& -1.36 &	-3.156 &	-2.676 &	-1.22 &	-1.62 \\
&$\epsilon_{p}^{B^\prime B^\prime}$& 4.15 &	2.42 &	1.9 & 3.24 &	2.8 \\
&$t_{ss}^{B^\prime B^\prime}$&	0.0 & 0.0 &	0.0 &	0.0 &	0.0 \\
B$^\prime$-B$^\prime$&$t_{sp\sigma}^{B^\prime B^\prime}$& 0.0 & 0.0 &	0.0 & 0.0 & 0.0 \\
&$t_{pp\sigma}^{B^\prime B^\prime}$& 0.0 &	0.0 &	0.0 &	0.0 &	0.0 \\
&$t_{pp\pi}^{B^\prime B^\prime}$& 0.0 & 0.0 &	0.0 &	0.0 &	0.0 \\
\hline
&$t_{ss\sigma}^{BB^\prime}$& -0.23 & -0.377 &	-0.303 &	-0.21 &	-0.22  \\
&$t_{sp\sigma}^{BB^\prime}$& 0.547 & 0.44 &	0.51 &	0.50 &	0.47 \\
B-B$^\prime$&$t_{pp\sigma}^{BB^\prime}$& 0.805 &	0.665 &	0.665 &	0.675 &	0.72 \\
&$t_{pp\pi}^{BB^\prime}$& 0.105 & 0.09 &	0.09 &	0.089 &	0.078 \\
\hline
SOC&$\lambda^B$& 0.52 &	0.16&	0.16&	0.46&	0.37 \\
&$\lambda^{B^\prime}$& 0.52 &	0.56&	0.17&	0.49&	0.17 \\
\hline
\end{tabular}
\label{T3}
\end{table*}
\renewcommand{\thetable}{S12}
\begin{table*}[h]
\centering
\caption {TB hopping parameters (in eV) for halide perovskites family.}
\begin{tabular}{ |c|c|c|c|c|c|} 
\hline
Cation- &	TB parameters &	$Cs_2NaInCl_6$  & $Cs_2NaBiCl_6$  & $Cs_2KInCl_6$  & $Cs_2KBiCl_6$ \\
cation & &	B = Na, B$^\prime$ = In  & B = Na, B$^\prime$ = Bi& B = K, B$^\prime$ = In  & B = K, B$^\prime$ = Bi\\
\hline
&$\epsilon_s^B$&  12.4 &	9 & 15 & 15 \\
&$\epsilon_{p}^B$& 13.5 &	10 &	15 & 15 \\
B-B&$t_{ss\sigma}^B$&	-- &	-- &	-- & -- \\
&$t_{sp\sigma}^B$&	-- &	-- &	-- & -- \\
&$t_{pp\sigma}^B$&	-- &	-- &	-- & -- \\
&$t_{pp\pi}^B$& -- &	-- &	-- &	-- \\
\hline
&$\epsilon_{s}^{B^\prime B^\prime}$&	5.8 &	-0.08 &	6.21 &	0.0\\
&$\epsilon_{p}^{B^\prime B^\prime}$&	9.64 &	5.20 &	9.35 &	5.12\\
&$t_{ss\sigma}^{B^\prime B^\prime}$&	-0.036 &	-0.034 &	0.010 &	0.01\\
B$^\prime$-B$^\prime$&$t_{sp\sigma}^{B^\prime B^\prime}$& 0.018 &	0.05 &	0.05 &	0.04 \\
&$t_{pp\sigma}^{B^\prime B^\prime}$&	-0.152 &	-0.08 &	-0.15 &	-0.01 \\
&$t_{pp\pi}^{B^\prime B^\prime}$& 0&	0.02 &	0 &	0\\
\hline
&$t_{ss\sigma}^{BB^\prime}$& 0.0 &  0.0 &	-0.33 &	-0.40\\
&$t_{sp\sigma}^{BB^\prime}$& 0.0 &	0.0 0 &	0.0 &	0.0\\
B-B$^\prime$&$t_{pp\sigma}^{BB^\prime}$& 0.0 & 0.0 &	0.0 &	0.0\\
&$t_{pp\pi}^{BB^\prime}$& 0.0 &	0.0 &	0.0 &	0.0\\
\hline
SOC&$\lambda^B$&	0.0&	0.0&	0.0&	0.0\\
&$\lambda^{B^\prime}$&	0.16&	0.52&	0.16 &	0.52\\
\hline
\end{tabular}
\label{T4}
\end{table*} 

\newpage
\pagebreak
\section{{B-MOP and tight-binding model for $\bf{Cs_2AgAsCl_6}$, $\bf{Cs_2AgGaCl_6}$, $\bf{Cs_2CuBiCl_6}$, and $\bf{Cs_2CuInCl_6}$}}

We have added four more compounds to our list of HDPs and performed the model Hamiltonian calculations for the same. These compounds are Cs$_2$AgAsCl$_6$, Cs$_2$AgGaCl$_6$, Cs$_2$CuBiCl$_6$, and Cs$_2$CuInCl$_6$. We have obtained the B-MOP for all these compounds and performed the tight-binding calculations. Upon careful examination, we deduce a consistent pattern wherein electronic characteristics remain largely unaltered, apart from the bandgap, when newly substituted cations belong to the same group within the periodic table. This phenomenon is exemplified by Cs$_2$AgB$^\prime$Cl$_6$, where B$^\prime$ corresponds to group 15 elements (As, Sb, Bi) within the periodic table. In each of these three compounds, the band structure consistently exhibits the qualities of an indirect bandgap, with the VBM situated at point X, and the CBM located at the high-symmetry L. Notably, as we turn B$^\prime$ from As to Bi within this framework, a remarkable trend unfolds. The mBJ bandgap experiences a progressive increase, escalating from 2.26 to 2.98 eV, while concurrently, the dispersion of the edge bands diminishes. Similar observations have been made for Cs$_2$AgB$^\prime$Cl$_6$, where B$^\prime$ corresponds to group 13 elements (Ga, In, Tl). The nature of bandgap remains direct with both VBM and CBM situated at $\Gamma$. The calculated MME also shows the same behavior with the parity-forbidden transition from $\Gamma$ to X for all three cases (see Fig. \ref{new_TB_bands}) however, the bandgap is maximum for B$^\prime$ = In. The reason behind it can be explained through the free atomic energy of B$^\prime$-$s$ orbitals. The TB fitting parameters for all four new compounds are given in Table \ref{T2}.

In another case, we altered B from Ag to Cu and conducted electronic structure calculations using various B$^\prime$ elements. The calculated band structures, B-MOP results, tight-binding models, MME, and parity values from symmetry analysis are illustrated in Fig. \ref{new_TB_bands}.
\renewcommand{\thefigure}{S2}
\begin{figure}
    \centering
    \includegraphics[scale=0.4]{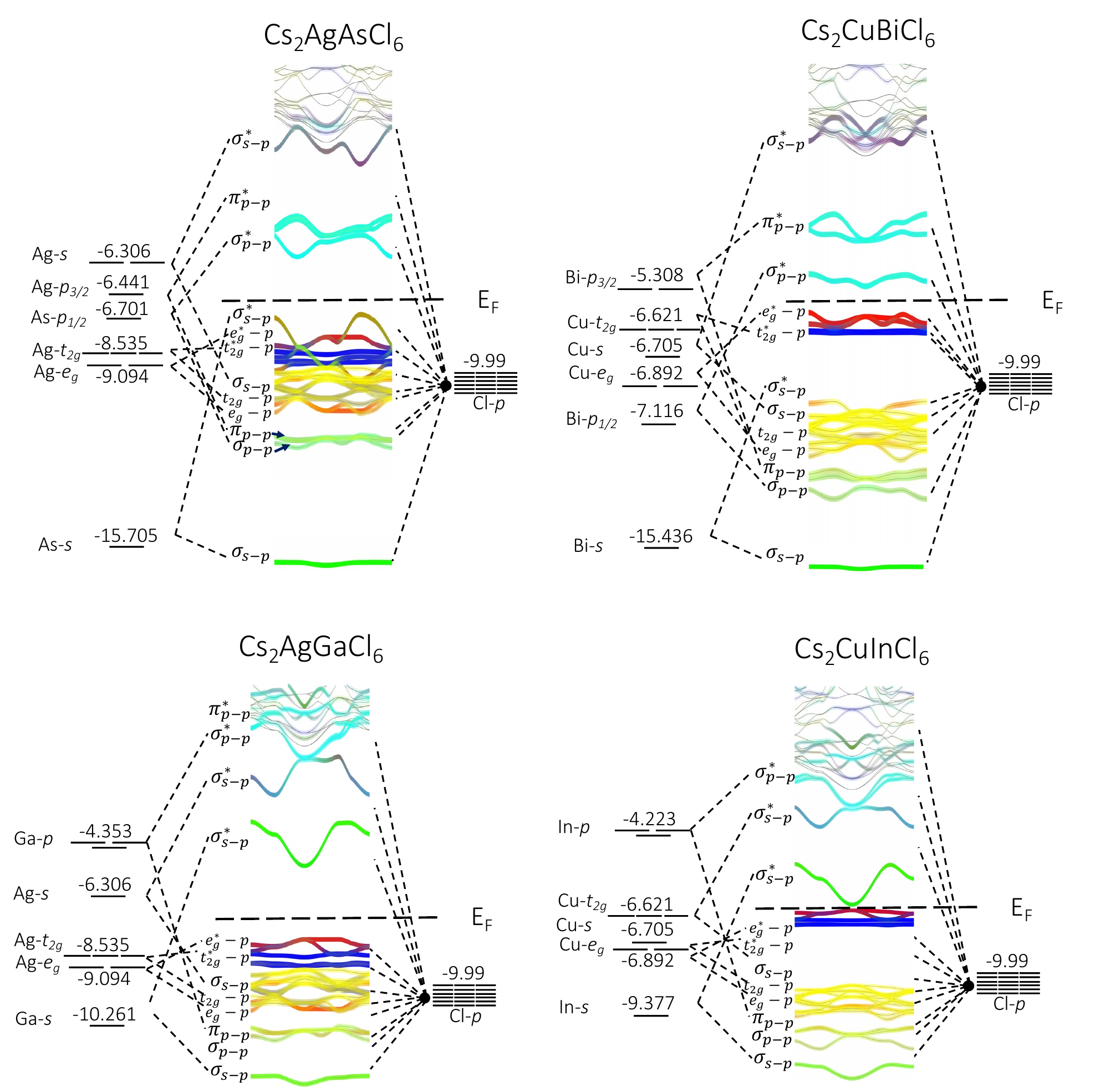}
    \caption{The band projected molecular orbital picture (B-MOP) of halide double perovskites as envisaged from the following molecular hybridizations: B(Ag, Cu)-{$s$, $d$}-Cl-$p$ and B$^\prime$(Ga, As, Bi, In)-{$s$, $p$}-Cl-$p$ atomic orbitals produces the bonding and antibonding orbitals along with the nonbonding Cl-$p$ orbitals. The free atomic energy levels are estimated from Hartree-Fock’s theory. }
    \label{new-BMOP}
\end{figure}
\renewcommand{\thefigure}{S3}
\begin{figure}
    \centering
    \includegraphics[scale=0.12]{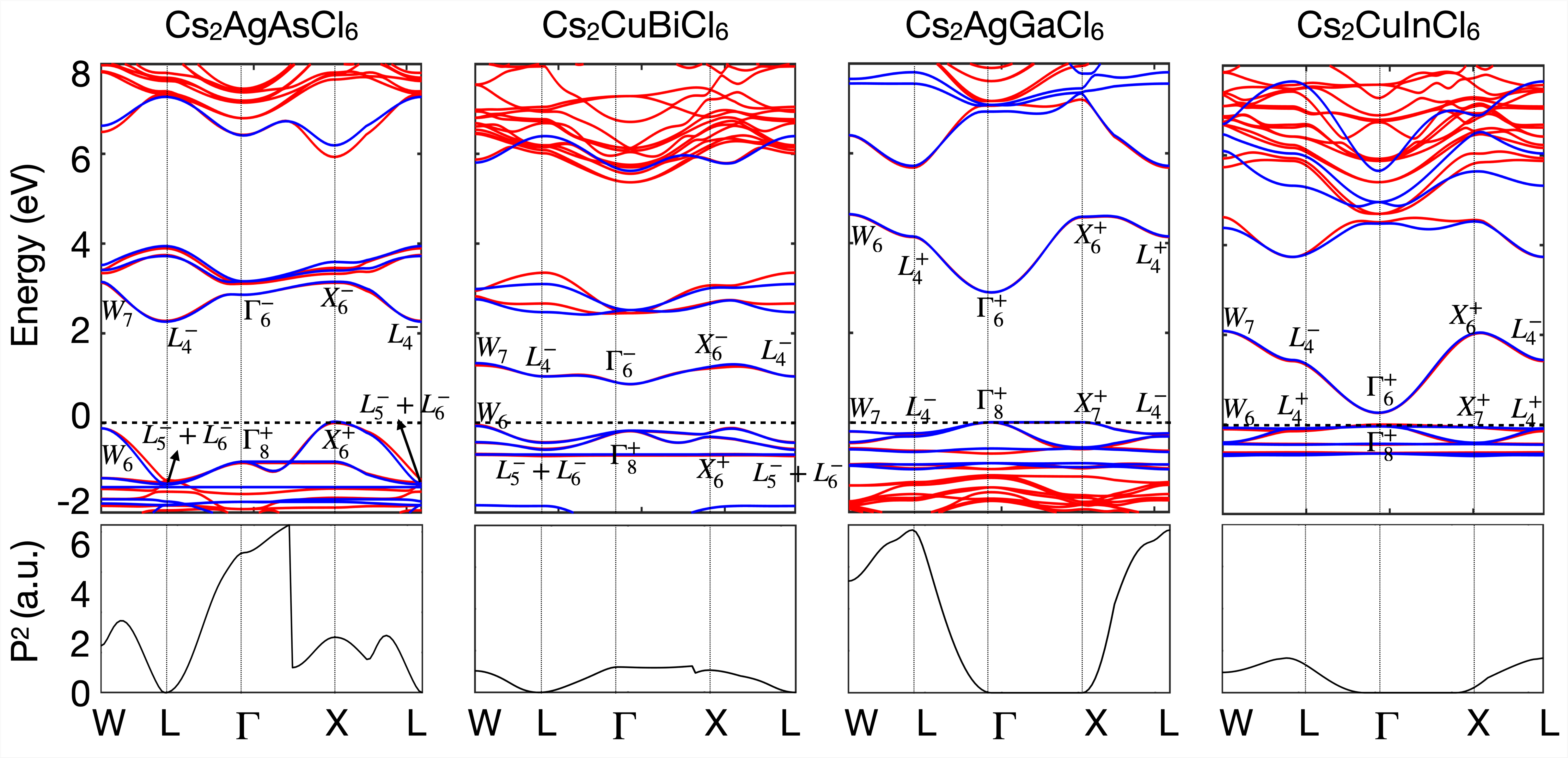}
    \caption{DFT (red) fitted tight-binding bands (blue) of Cs$_2$AgAsCl$_6$, Cs$_2$AgGaCl$_6$, Cs$_2$CuBiCl$_6$, and Cs$_2$CuInCl$_6$. The squared momentum matrix elements (P$^2$) corresponding to the valence band edge (VBE) to conduction band edge (CBE) transition are also shown for each of the compounds. The parity of the VBE and CBE are mentioned using Koster notations.}
    \label{new_TB_bands}
\end{figure}

\newpage
\pagebreak

\section{Mathematical formulation to calculate optical properties}
We will start from the Fermi Golden rule and then derive the expressions for the momentum matrix element (MME).\\

\textbf{Fermi Golden rule:} The Fermi Golden rule states that transition probability per unit time from initial state  $\ket{i}$  to a set of final states  $\ket{f}$ is given to first-order approximation:
\begin{equation}
    W_{i\rightarrow f} = \frac{2\pi}{\hbar} |\bra{\textbf{k}, \beta}\vec{P}\ket{\textbf{k}, \beta^\prime}|^2 \rho(E_F)
\end{equation}
where $\bra{\textbf{k}, \beta}\vec{P}\ket{\textbf{k}, \beta^\prime}$ is matrix element of perturbation H$^\prime$ between final and initial states, and $\rho(E_f)$ is density of states at energy E$_F$. The density of states for transition between states $E_v$ and $E_c$ with incident photon energy $\hbar \omega$, can be approximated to $\delta(E_c - E_v - \hbar \omega)$.   \par
As a starting point to derive the momentum matrix element (MME), let us consider $n^{th}$  tight-binding eigenstate formed with a linear combination of atomic orbitals, 
\begin{equation}
    \ket{n,\textbf{k}} = \sum_\alpha c_{n \alpha} (\textbf{k}) \ket{\alpha, \textbf{k}} 
\end{equation}
where $\ket{\alpha, \textbf{k}}$ is an atomic orbital eigenstate in $k$ space, $\alpha$ is the atomic orbital, and $k$ is momentum. In periodic systems, the state $\ket{\alpha, \textbf{k}}$ can be defined as the Bloch sum of localized orbitals
\begin{equation}
    \ket{\alpha,\textbf{k}} = \frac{1}{\sqrt{N}}\sum_\textbf{R} e^{i\textbf{k} \cdot \textbf{R}} \ket{\alpha, \textbf{R}} .
\end{equation}
Where \textbf{R} is lattice vector and N is number of lattice sites. The system Hamiltonian $H$ is then diagonalized within eigenstate to obtain the energy eigenvalue $E_{n \textbf{k}}$ of the system.
\begin{equation}
     H(k) \ket{n, \textbf{k}} = E_{nk} \ket{n, \textbf{k}} \\
\end{equation}
The connection between position and momentum operator is derived through commutator relation:
\begin{equation}
\begin{split}
    [H, \textbf{r}] &= \left[\frac{p^2}{2m_e} +V_0(\textbf{r}) , \textbf{r} \right] \\
    [H, x] &= \frac{1}{2m_e}([p_x[p_x,x]+[p_x,x]p_x])\\
    [H, x] &= \frac{-i\hbar}{m_e}p_x\\
\end{split}
\end{equation}
Similarly, one can obtain $[H, y] = (-i\hbar/m_e)p_y$ and $[H, z] = (-i\hbar/m_e)p_z$. Hence,
\begin{equation}
    [H, \textbf{r}] = \frac{-i\hbar}{m_e}\textbf{p}
\end{equation}
or,
\begin{equation}
    \textbf{p} = \frac{im_e}{\hbar}[H, \textbf{r}]
\end{equation}
\newpage
\begin{equation}
\begin{split}
    & \bra{n,k}\textbf{p}\ket{m,k} = \bra{n,k}\frac{im_e}{\hbar}[H, \textbf{r}]\ket{m,k}\\
    &= \frac{im_e}{\hbar}\bra{n,k}H\textbf{r}-\textbf{r}H\ket{m,k}\\
    &= \frac{im_e}{\hbar} \sum_\beta c_{n\beta}^*(k)\frac{1}{\sqrt{N}}\sum_{R^\prime} \bra{\beta, R^\prime}e^{-i\vec{k}\cdot\vec{R^\prime}}(H\textbf{r}-\textbf{r}H)\sum_\alpha c_{m\alpha}(k)\sum_R\frac{1}{\sqrt{N}}\ket{\alpha, R}e^{i\vec{k}\cdot\vec{R}}\\
    &= \frac{im_e}{\hbar N}\sum_{\alpha\beta}c_{n\beta}^*(k)c_{m\alpha}(k) \sum_{RR^\prime}e^{1k(R-R^\prime)}\bra{\beta R^\prime}H\textbf{r}-\textbf{r}H\ket{\alpha R}\\
    &= \frac{im_e}{\hbar N}\sum_{\alpha\beta}c_{n\beta}^*(k)c_{m\alpha}(k) \sum_{RR^\prime}e^{ik(R-R^\prime)} (\bra{\beta R^\prime}H\textbf{r}\ket{\alpha R} - \bra{\beta R^\prime}\textbf{r}H\ket{\alpha R}) \\
    &= \frac{im_e}{\hbar N}\sum_{\alpha\beta}c_{n\beta}^*(k)c_{m\alpha}(k) \sum_{RR^\prime}e^{ik(R-R^\prime)} (\bra{\beta R^\prime}H \sum_\gamma \ket{\gamma R}\bra{\gamma R} \textbf{r}\ket{\alpha R} \\
    & \hspace{6cm} - \bra{\beta R^\prime}\textbf{r} \sum_\gamma \ket{\gamma R^\prime}\bra{\gamma R^\prime} H\ket{\alpha R}) \\
    &= \frac{im_e}{\hbar N}\sum_{\alpha\beta}c_{n\beta}^*(k)c_{m\alpha}(k) \sum_{RR^\prime}e^{ik(R-R^\prime)} (\sum_\gamma \bra{\beta R^\prime}H \ket{\gamma R}\ (R\delta_{\gamma\alpha}+d_{\alpha \gamma}) - \\
    & \hspace{6cm} \sum_\gamma\bra{\gamma R^\prime}H \ket{\alpha R} (R^\prime \delta_{\gamma \beta} + d_{\gamma\beta})) \\
    &= \frac{im_e}{\hbar N}\sum_{\alpha\beta}c_{n\beta}^*(k)c_{m\alpha}(k) \sum_{RR^\prime}e^{ik(R-R^\prime)} \sum_\gamma(\bra{\beta R^\prime}H\ket{\gamma R}R\delta_{\gamma\alpha}  - \bra{\gamma R^\prime}H\ket{\alpha R} R^\prime \delta_{\gamma \beta}\\
    & \hspace{6cm}  + d_{\alpha \gamma}\bra{\beta R^\prime}H\ket{\gamma R} - d_{\gamma\beta}\bra{\gamma R^\prime}H\ket{\alpha R})\\
    &= \frac{im_e}{\hbar N}\sum_{\alpha\beta}c_{n\beta}^*(k)c_{m\alpha}(k) \sum_{RR^\prime}e^{ik(R-R^\prime)} (\bra{\beta R^\prime}H\ket{\alpha R}(R - R^\prime) + \\ 
    & \hspace{6cm}  \sum_\gamma (\bra{\beta R^\prime}H\ket{\gamma R} d_{\alpha \gamma} - \bra{\gamma R^\prime}H\ket{\alpha R}d_{\gamma \beta}))\\
    &= \frac{im_e}{\hbar N}\sum_{\alpha\beta}c_{n\beta}^*(k)c_{m\alpha}(k) \sum_{RR^\prime}e^{ik(R-R^\prime)} (R-R^\prime) \bra{\beta R^\prime}H\ket{\alpha R} + \frac{im_e}{\hbar N}\sum_{\alpha\beta}c_{n\beta}^*(k)c_{m\alpha}(k) \\
    & \hspace{6cm} \sum_{RR^\prime}e^{ik(R-R^\prime)} \sum_\gamma (\bra{\beta R^\prime}H\ket{\gamma R}d_{\alpha \gamma} - \bra{\gamma R^\prime}H\ket{\alpha R} d_{\gamma \beta})\\
    &= \frac{im_e}{\hbar N}\sum_{\alpha\beta}c_{n\beta}^*(k)c_{m\alpha}(k) (-i\nabla_k)\sum_R\sum_{R^\prime} e^{ikR}e^{-ikR^\prime} \bra{\beta R^\prime}H\ket{\alpha R} + \frac{im_e}{\hbar N}\sum_{\alpha\beta}c_{n\beta}^*(k)c_{m\alpha}(k) \\
    & \hspace{6cm} \sum_R\sum_{R^\prime} e^{ikR}e^{-ikR^\prime} \sum_\gamma (\bra{\beta R^\prime}H\ket{\gamma R}d_{\alpha \gamma} - \bra{\gamma R^\prime}H\ket{\alpha R} d_{\gamma \beta}) \\
    &= \frac{m_e}{\hbar}\sum_{\alpha\beta}c_{n\beta}^*(k)c_{m\alpha}(k) \nabla_k \bra{\beta k}H\ket{\alpha k} + \frac{im_e}{\hbar}\sum_{\alpha\beta}c_{n\beta}^*(k)c_{m\alpha}(k) \sum_\gamma (\bra{\beta k}H\ket{\gamma k}d_{\alpha \gamma} - \\
    & \hspace{6cm} \bra{\gamma k}H\ket{\alpha k} d_{\gamma \beta}) \\
    &= \frac{m_e}{\hbar}\sum_{\alpha\beta}c_{n\beta}^*(k)c_{m\alpha}(k) \nabla_k \bra{\beta k}H\ket{\alpha k} + \frac{im_e}{\hbar}\sum_{\alpha\beta}c_{n\beta}^*(k)c_{m\alpha}(k) (E_{\beta k} - E_{\alpha k})d_{\alpha \beta} \\
    &= \frac{m_e}{\hbar}\sum_{\alpha\beta}c_{n\beta}^*(k)c_{m\alpha}(k) \nabla_k \bra{\beta k}H\ket{\alpha k} +  (E_{n, k} - E_{m, k})\frac{im_e}{\hbar}\sum_{\alpha\beta}c_{n\beta}^*(k)c_{m\alpha}(k)d_{\alpha \beta} \\
\end{split}
\end{equation}
Recall that the energy eigenstate $\ket{n, k}$ can be written in the Bloch form:
\begin{equation}
    \ket{n,\textbf{k}} = e^{i\textbf{k}\cdot\textbf{r}} \ket{u_{n\textbf{k}}}, 
\end{equation}
where the cell-periodic wave function $\ket{u_{n\textbf{k}}}$ satisfies
\begin{equation}
    \hat{H}_u(\textbf{k})\ket{u_{n\textbf{k}}} = \epsilon_{n\textbf{k}}\ket{u_{n\textbf{k}}}
\end{equation}
Thanks to $\ket{u_{n\textbf{k}}}$, the MME for periodic Block functions $ u_{\textbf{k}\beta}$ $ e^{i\textbf{k}\cdot \textbf{R}}$ are obtained as follows:
\begin{equation}
    \bra{\textbf{k}, \beta}\vec{P}\ket{\textbf{k}, \beta^\prime} = \frac{m_e}{\hbar}\bra{u_{\textbf{k}\beta}}\frac{\partial H(\textbf{k})}{\partial \textbf{k}}\ket{u_{\textbf{k}\beta^\prime}}.
\end{equation}
In the case of optical transition, the transition from the top valence to the bottom conduction band is considered. Therefore, the component of relevant MME  can be expressed as follows,
\begin{equation}
    (P_{v,c})_{x,y,z} = \frac{m_e}{\hbar}\sum_{\beta,\beta^\prime} u_{\textbf{k}\beta,c} \frac{\partial H_{\beta\beta^\prime}}{\partial k_{x,y,z}} u_{\textbf{k}\beta^\prime,v}.
\end{equation}
Here, $u_{\textbf{k}\beta,c}$ and $u_{\textbf{k}\beta^\prime,v}$ respectively represent the eigenvectors associated with the energy eigenvalues $E_v$ and $E_c$. The real and imaginary parts of the dielectric constant can be calculated as: 
\begin{equation}
\begin{split}
    \epsilon_1(\omega) &= 1+\frac{2}{\pi}C \int_0^{\inf}  \frac{\omega^\prime\epsilon_2(\omega^\prime)}{\omega^{\prime 2}-\omega^2}d\omega^\prime, \\\\ \nonumber
    \epsilon_2(\omega) &= \frac{e^2\hbar^2}{\pi m_e^2\omega^2}\sum_{v,c} \int_{BZ}d^3k|P_{v,c}|^2\times \delta(E_c(\textbf{k})-E_v(\textbf{k})-\hbar\omega).
\end{split}
\end{equation}
and
\begin{equation}
    \epsilon(\omega) = \epsilon_1(\omega)+i\epsilon_2(\omega)
\end{equation}
Here, C is the Cauchy principal value of integral; $e$ and $m_e$,  respectively, are charge and mass of an electron. P$_{v,c}$ in $\epsilon_2(\omega)$ is MME corresponding to a transition from the valence band at energy $E_v$ to the conduction band at energy $E_c$. The Dirac-delta function switches on MME contribution when a transition occurs from one state to another. E$_c$, E$_v$ and eigenvectors are taken from our developed TB model. The optical absorption coefficient $\alpha(\omega)$ of a material is determined by its frequency-dependent dielectric constant:
\begin{equation}
       \alpha(\omega) = \omega \sqrt{\frac{-\epsilon_1(\omega) + \sqrt{\epsilon_1^2(\omega) + \epsilon_2^2(\omega)}}{2}}
\end{equation} 
\section{TB band structure of cation doped halide double perovskites}
From the hopping interaction strength, we observed that electron's hoppings are confined to 2$^{nd}$ neighbor B-B$^\prime$ interactions. Therefore TB fitting parameters obtained for the pristine unit cell can be further used to calculate the electronic properties of cation-doped double perovskites. Here, we have designed the model for Cs$_2$AgIn$_{1-x}$Bi$_x$Cl$_6$ and Cs$_2$Na$_{1-x}$Ag$_x$InCl$_6$ at $x$ = 0.25, 0.50, and 0.75. To obtain TB band structure, we have designed four formula supercell, from rhombohedral unit cell to cubic supercell as shown in Fig. 2  of the main text, which contains eight metal cation sites in it. In the supercell crystal structure, four sites are filled with Ag cations and rest sites have been distributed between In and Bi cations depending on the value of $x$ in Cs$_2$AgIn$_{1-x}$Bi$_x$Cl$_6$. Similarly, the Hamiltonian for Cs$_2$Na$_{1-x}$Ag$_x$InCl$_6$ is also designed. The DFT band structure for both Cs$_2$AgIn$_{1-x}$Bi$_x$Cl$_6$ and Cs$_2$Na$_{1-x}$Ag$_x$InCl$_6$ compounds, calculated along high symmetry k-path shown in Fig. 2 (c) in the main text, are shown in Fig. \ref{fig:S1}.
\renewcommand{\thefigure}{S4}
\begin{figure}[ht]
    \centering
    \includegraphics[scale=0.14]{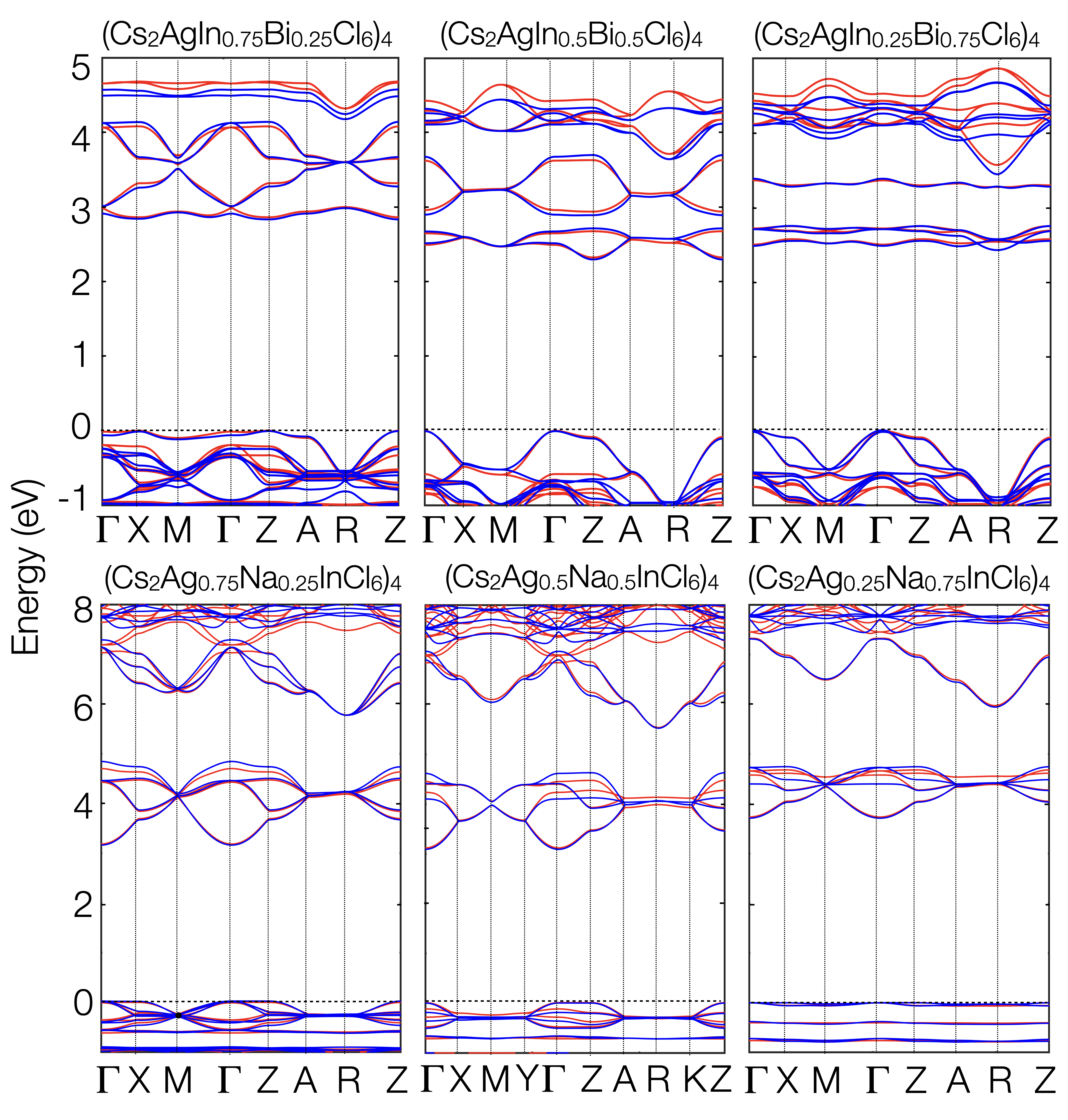}
    \caption{DFT (red) fitted TB (red) bands of cation doped halide perovskites Cs$_2$AgIn$_{1-x}$Bi$_{x}$Cl$_6$ and Cs$_2$Ag$_{1-x}$Na$_x$InCl$_6$.}
    \label{fig:S1}
\end{figure}
\section{TB fitted onsite energy of different orbitals for cation doped HDP}
The TB hopping parameters, while cation doping, are fixed (do not vary with $x$) and taken from previously fitted pristine compounds. However, the onsite energies vary as a function of $x$ and are plotted in Fig. \ref{fig:S2}. The polynomial expressions of onsite energies of  Cs$_2$AgIn$_{1-x}$Bi$_x$Cl$_6$ and Cs$_2$Ag$_{1-x}$Na$_x$InCl$_6$ are provided in Eq. 23 and 24.  Since X-\{$p$\} orbitals are omitted from the model Hamiltonian, and M-X interactions are downfolded to effective onsite energies\cite{RaviJCP}. Only onsite energies of Ag-\{$s$, $d$\}, In-\{$s$, $p$\}, and Bi-\{$s$, $p$\} are varying as a function of $x$.\\

In Fig. \ref{fig:S2}, we have shown the variation of different onsite interaction parameters for these two cation doped HDPs as a function of $x$.
The polynomial expressions for onsite energy vs. doping concentration $x$ for Cs$_2$AgIn$_{1-x}$Bi$_x$Cl$_6$ are; 
\begin{eqnarray}
    \epsilon_s^{Ag}(x) & = & -5.01x^4+19.25x^3-16.25x^2+2.41x+7.15,\\ \nonumber
    \epsilon_{eg}^{Ag}(x)& = & -8.70x^4+18.46x^3-10.34x^2+0.28x-0.08,\\ \nonumber
    \epsilon_{t2g}^{Ag}(x)& = & -6.08x^4+15.34x^3-9.46x^2+0.14x-0.82,\\ \nonumber
    \epsilon_s^{In}(x) & = & 4.60x^3-3.58x^2-3.32x+5.58,\\ \nonumber
    \epsilon_p^{In}(x) & = & 5.97x^3-8.48x^2+1.90x+7.94,\\ \nonumber
    \epsilon_s^{Bi}(x) & = & -5.97x^3+6.64x^2-3.20x-0.66,\\ \nonumber
    \epsilon_p^{Bi}(x) & = & 0.53x^3+2.64x^2-2.09x+4.04.\\ \nonumber
\end{eqnarray}
Similarly, for Cs$_2$Ag$_{1-x}$Na$_x$InCl$_6$, the expressions are:
\begin{eqnarray}
    \epsilon_s^{Ag}(x) & = & 0.1783x^3-1.125x^2+1.907x+6.19,\\ \nonumber
    \epsilon_{eg}^{Ag}(x)& = & 0.045x^3-0.285x^2+0.554x-0.395,\\ \nonumber
    \epsilon_{t2g}^{Ag}(x)& = & 0.063x^3-0.410x^2+0.806x-1.280,\\ \nonumber
    \epsilon_s^{In}(x) & = & 0.075x^3-0.378x^2+0.267x+4.632,\\ \nonumber
    \epsilon_p^{In}(x) & = & 0.135x^3-0.948x^2+1.917x+6.872,\\ \nonumber
\end{eqnarray}
\renewcommand{\thefigure}{S5}
\begin{figure}[ht]
    \centering
    \includegraphics[scale=0.25]{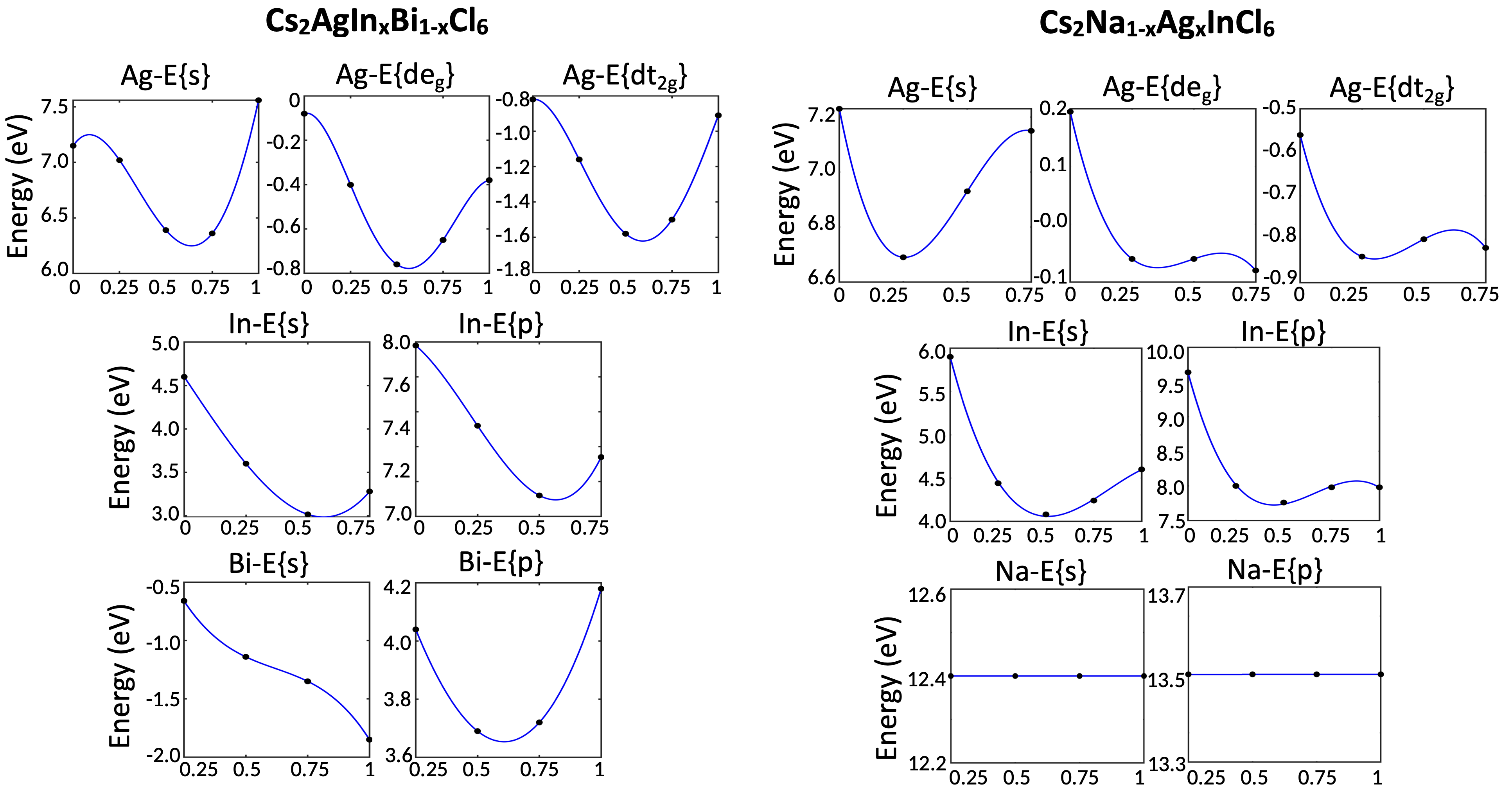}
    \caption{Various onsite energies of molecular orbitals as a function of $x$ in Cs$_2$AgIn$_x$Bi$_{1-x}$Cl$_6$ (left) and Cs$_2$Na$_{1-x}$Ag$_x$InCl$_6$ (right). The black circular dots represent the exact values obtained from curve-fitting, and the blue curve corresponds to the polynomial fitting of the black dots.}
    \label{fig:S2}
\end{figure}
\renewcommand{\thefigure}{S6}
\begin{figure}
    \centering
    \includegraphics[scale=0.22]{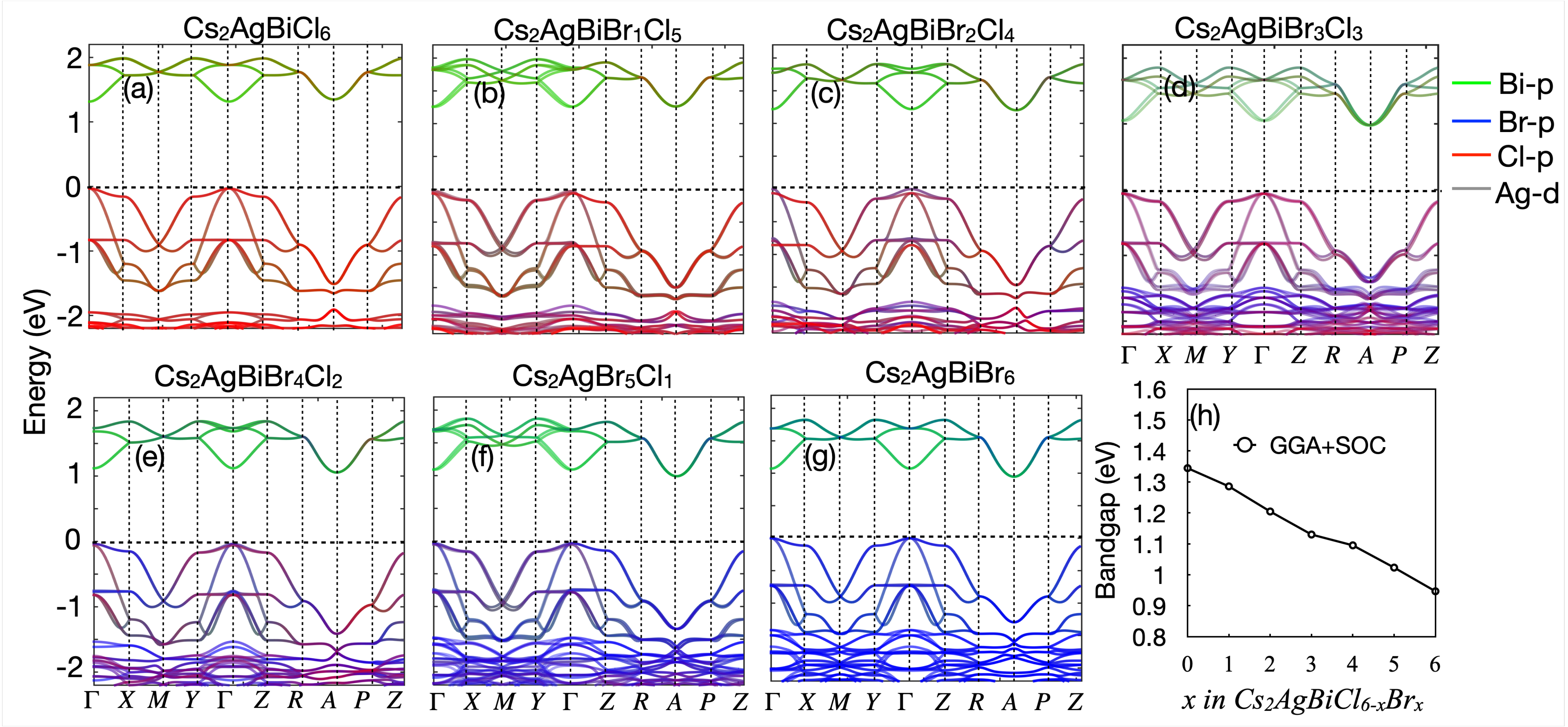}
    \caption{The orbital resolved band structured of Cs$_2$AgBiCl$_{6-x}$Br$_x$ as a function of $x$. The change in the bandgap from Br to Cl is approximately to the free atomic energy difference between Cl and Br ($\sim$0.5 eV). }
    \label{Cl-Br}
\end{figure}

\section{Discussion on Bromine doping in Chlorine based HDP}
As stated in our B-MOP, the positions of the bands in the energy spectrum are obtained from the molecular hybridization of (B, B$^\prime$)-\{$s$,$p$/$s$,$d$\} -- X-$p$ atomic orbitals. For example, in the case of Cs$_2$AgBiBr$_6$ and Cs$_2$AgBiCl$_6$, the prominent interactions that define the valence band spectrum and conduction band spectrum are Ag-$d$ -- Br/Cl-$p$ and Bi-$p$ -- Br/Cl-$p$, respectively. We find that the Br-$p$ energy levels are comparable to that of Ag-$d$ energy levels while the Cl-$p$ energy levels lie $\sim$ 0.5 eV below (9.96 eV for Cl-$p$ and 9.46 eV for Br-$p$ orbitals). Therefore, the Ag-$d$ — Br-$p$ interaction is stronger to push the antibonding $e_g^*$ band above for the Br compound as compared to that of the Cl compound. Since there is a large onsite energy difference between Bi-$p$ and Br/Cl-$p$ orbitals ($\epsilon_{(Bi-p)}-\epsilon_{(Br/Cl-p)}$), the positioning of the conduction band minimum ($\sigma_{(p-p)}^*$) is less affected by ($\epsilon_{(BrCl-p)}$). Thus, we believe that change in the bandgap is mainly due to the shift in the position of the valence band while changing Cl to Br.


To illustrate the bandgap variation as a function of halide anion, in Fig. \ref{Cl-Br}, we have plotted the orbital resolved band structures of Cs$_2$AgBiCl$_{6-x}$Br$_x$ and bandgap as a function of $x$. The band structures are calculated considering the computationally less expensive GGA functional, as the objective is to examine the trend. The  replacement of the Br anion from Cl anion gradually reduces the GGA bandgap of the compound from 1.34 eV for Cs$_2$AgBiCl$_{6}$ to 0.94 eV Cs$_2$AgBiBr$_{6}$. As depicted in Fig. \ref{Cl-Br}, we have noted that the orbital contribution of the edge bands remains consistent between Br and Cl. Additionally, as expected, we note that the contributions of Cl-$p$ orbitals (red) and Br-$p$ orbitals (blue) are more in the valence band spectrum in comparison to the conduction band spectrum. The net reduction in the bandgap is $\approx$ 0.45 eV, which is roughly the same order as the onsite energy difference between Cl-$p$ and Br-$p$ orbitals.  \par


\end{widetext}

\end{document}